\documentclass{article}
\usepackage{graphicx}
\usepackage{fullpage}

\usepackage{xspace}
\usepackage{url}
\usepackage{xcolor}
\usepackage{subcaption}

\newcommand{\SM}{Supplementary Material\xspace}
\newcommand{\sw}{{\em spaced word}\xspace}

\newcommand{\kmer}{$k$-mer\xspace}
\newcommand{\kmers}{$k$-mers\xspace}
\newcommand{\ignore}[1]{}

\usepackage{listings}
\newcommand{\vir}[1]{``#1''}

\title{Alignment-free Genomic Analysis via a Big Data Spark Platform}

\author{
Umberto Ferraro Petrillo\thanks{Dipartimento di Scienze Statistiche, Universit\`{a} di Roma - La Sapienza, Rome, 00185, Italy} \hspace{0.05mm} \thanks{To whom correspondence should be addressed.} \and 
Francesco Palini\footnotemark[1] \and 
Giuseppe Cattaneo\thanks{Dipartimento di Informatica, Universit\`{a} di Salerno, Fisciano (SA), 84084, Italy} \hspace{0.05mm} \thanks{Those two authors contributed equally to the research.} \and
Raffaele Giancarlo\thanks{Dipartimento  di Matematica ed Informatica, Universit\`{a} di Palermo, Palermo, 90133, Italy} \hspace{0.05mm} \footnotemark[4]
}

\date{}

\begin{document}

\maketitle

\begin{abstract}
\textbf{Motivation:} Alignment-free distance and similarity functions (AF functions, for short) are a well established alternative to two and multiple sequence alignments for many genomic, metagenomic and epigenomic tasks. Due to data-intensive applications, the computation of AF functions is a Big Data problem, with   the recent Literature indicating  that the development of  fast and scalable algorithms computing AF functions is a high-priority task. Somewhat surprisingly, despite the increasing popularity of  Big Data technologies in Computational Biology, the development of a Big Data platform  for those tasks has not been pursued, possibly due to its complexity.\\
\textbf{Results:} We fill this important gap by introducing FADE,  the first  extensible,  efficient and scalable Spark platform for Alignment-free genomic analysis.  It supports natively eighteen  of the best performing AF functions coming out of a recent hallmark benchmarking study. 
FADE development and potential impact comprises novel aspects of interest. Namely,  (a) a considerable effort of distributed algorithms, the most tangible result being  a much faster execution time of reference  methods like MASH and FSWM; (b)  a  software design that makes FADE  user-friendly and easily extendable by Spark non-specialists; (c) its ability to support data- and compute-intensive tasks. About this, we provide a novel and much needed analysis of how informative and robust AF functions are, in terms of the statistical significance  of their output. 
Our findings naturally extend the ones  of the highly  regarded benchmarking study, since the functions that can really be used are reduced to a handful of the eighteen included in FADE.\\
\textbf{Contact:} umberto.ferraro@uniroma1.it
\end{abstract}

\section{Introduction}

Alignment-free distance and similarity functions  (AF functions, for short) have been introduced as an alternative to traditional alignment-based methods, e.g, \cite{altschul1990basic,smith1981identification}, in order to assess how similar each pair of sequences in a collection are  to each other. 
By now, their use has been widely investigated for sequence analysis in genomics \cite{Zielezinski2019}, metagenomics \cite{Benoit16}, and epigenomics \cite{Giancarlo2015, Giancarlo18}.  The pros/cons of AF functions with respect to their alignment counterparts is well presented in \cite{Zielezinski2019}.  One of the key pro features highlighted in that study is that their implementations  offer data scalability, opportunities that alignment methods lack. 
Taking into account the throughput of HTS technologies, another compelling case on how  important is to design and implement fast and scalable algorithms for the computation of AF functions is presented in \cite{Ondov2016}. It is of interest here to notice that the mentioned two studies clearly indicate that the computation of AF functions is now a Big Data problem. 

As such, it needs algorithmic solutions that use Big Data Technologies to grant efficiency and scalability as a function of the available computational resources and of the amount of data to process. Somewhat surprisingly, although those technologies are finding more and more use in  Computational Biology \cite{Cattaneo19,mushtaq2015cluster}, and Cloud Storage and Computing is the future  of genomic data \cite{Kahn11022011}, only a few studies in that area are available, concentrating on AF methods for 
informational and linguistic analysis of genomic sequences \cite{cattaneo2016effective}. 
Yet, an effective Big Data platform supporting both the computation of AF functions and based on a pervasive  Big Data framework such as Spark, would place AF sequence comparison  at a peer with other important domains in Data Science (see, e.g., \cite{gonzalez2012powergraph}) that have such platforms,  as well as  contribute to  increase their usage  in the Life Sciences. 

\begin{figure}[t]
	\centering
	\includegraphics[width=0.9\textwidth]{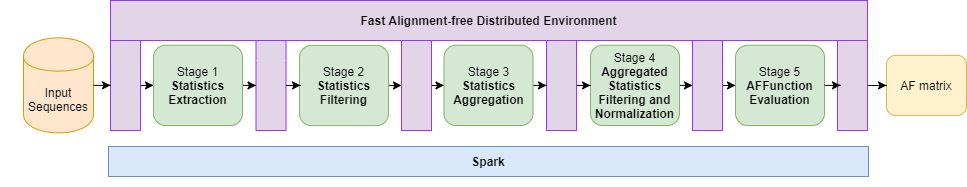}
	\caption{A layout of the logical architecture of the basic pipeline for the fast computation of AF functions. }
	\label{main:fig:fade}
\end{figure}

Our first contribution is to address this acute need. Indeed, we propose FADE,  the first  extensible and scalable Spark platform for the effective computation of AF functions. In its starting configuration, it offers eighteen implementations of highly performing AF functions according to results in \cite{Zielezinski2019}.   The basic pipeline of our platform consists of five (possibly optional) stages, as outlined in Figure \ref{main:fig:fade}.  
Its key features are as follows.

 \begin{itemize}
 	
 	\item{\textbf{FADE is User-Friendly and Easily Extensible}} 
 	FADE comes as a ready-to-use Spark application that can be easily executed without writing any line of code. If needed, its standard processing can be customized by just editing a provided reference configuration file, so as to choose from an included library which statistics to extract and which AF functions to evaluate. Some examples of configuration files are reported in Section 3.1 of the \SM.  The user can add support for a target statistic or a target AF function not originally included in the library, by writing the corresponding code using one of the provided class templates. This part is outlined in Section \ref{subsubsec:basicpipeline}. 
 	
\item{\textbf{FADE is Scalable and Efficient.}}  
In order to assess FADE ability to profitably support the implementation of AF functions, we have compared two of the most  prominent ones in the shared memory category, i.e., Mash \cite{Ondov2016}  and FSWM \cite{Leimester17}, vs their respective implementations supported by FADE (denoted with the prefix FADE). Being those latter based on a distributed framework, they are able to take advantage of the much higher number of processing cores available, achieving performances that are much better than those of shared memory tools and that  scale as a function of the processing units available. It is to be noted that, given the wide range of application scenarios for AF sequence comparison, varying from the examination of a large collection of very small reads, up to the  comparison of a few huge genomic sequences, a flexible  workload assignment is fundamental  to grant scalability and speed. To this end, FADE provides, and we have experimented with,  three workload choices. Transparently from the user, each of the three is associated to a transformation of the logical basic pipeline into a suitable \vir{run time} software pipeline, which is then executed. The Partial Aggregation strategy results to be the most appropriate, with varying workloads.
All the results, and the corresponding experiments, regarding this part  are presented in Section \ref{sec:intuition3}. 
 \end{itemize}

Our second contribution, introduced in Section \ref{subsec:reliability} and detailed in Section \ref{subsec:reliabilityexp}, shows  the ability of FADE to tackle data and compute intensive tasks and it is of interest in its own right. Indeed, we use novel ideas, that boil down to very costly Monte Carlo Simulations,  to gain  insights into the properties of AF functions, going a step further in the direction indicated  by the mentioned benchmarking study. The end result is new guidelines on the use of AF functions in day-to-day genomic analysis tasks. 

To this end, we consider reliability and robustness of AF functions in terms of the statistical significance of the distance/similarity matrices they   produce on benchmark datasets. In details, we consider a p-value obtained via a Monte Carlo simulation with the Null Hypothesis being that the values obtained by a given AF function on two biological sequences is no better than the value obtained on two random sequences. A value is significant when the Null Hypothesis is rejected. We account for repeated tests by applying  Bonferroni Correction. Our key findings are the following.

\begin{itemize}
	
	\item{\textbf{A Novel Class of AF functions: Consistently Significant.}}  
Across benchmark datasets from \cite{Zielezinski2019},  only a handful of the eighteen AF functions we have considered provide distance/similarity matrices for which the Null Hypothesis   is rejected  for the vast majority of their entries.  We refer to those functions as \emph{consistently significant}, since they behave consistently well,  statistically, irrespective of the dataset, and it is well known that statistical relevance is a good indication of biological relevance \cite{DidFr02,  Giancarlo2008,speed96}. They are  members of the D2 family \cite{kai2013} (D2 and D2*), and FSWM \cite{Leimester17}. Since the D2 family has been studied extensively from the statistical point of view, our results confirm that it is an excellent choice. For FSWM, this analysis is completely new. This part is in Section \ref{subsec:sigtest}.


\item{\textbf{Sensitivity to Noise of Consistently Significant AF Functions.}}  
In order for the consistently significant AF functions to be reliably useful for \vir{everyday analysis}, it is also important to assess how sensitive they are to the presence of \vir{noise}. That is, how the performance of an AF matrix varies as a function of the number of its entries that are not statistically significant. To the best of our knowledge, this sensitivity analysis  is completely new and the findings are very informative. It is described in Section \ref{subsec:sensitivity}. Based on those results, a small amount of "noise" is enough to have a significant reduction in performance. Therefore, the operational range of those functions is limited to AF matrices that pass at least 95\% of the statistical significance test. Interestingly, since we used philogenetic tree construction for this sensitivity analysis, we find that UPGMA  \cite{sneath1973} is better than the Neighbor Joining method \cite{saitou1987}) (NJ, for short), in dealing with small amounts of \vir{noisy} entries in the AF matrices.  This aspect of those two methods is new.
	
\end{itemize}

In the reminder of this paper, we assume that the reader is familiar with Apache Spark \cite{spark}. For the convenience of the reader, and due to space limitations, a short primer is reported in Section 1 of the \SM. 

\nopagebreak
 
\section{Methods}

\subsection{A Selection of Alignment-free Distances and Similarity Functions}\label{sec:funct}

The  two most popular  classes of  AF functions   are those based on {\em \kmer statistics} and  on {\em word matches}, sometimes also referred to as {\em micro alignments}.  For this research, we consider those two classes, providing also motivation for our choices.

An AF function  in the first class   is   based on {\em \kmer statistics}  (or { \em histogram statistics} \cite{luczak2017survey}). and  it  can be used to analyze  a set of sequences  as follows. For each sequence in the set, the contiguous subwords of length $k$ therein contained (i.e., \kmers) with their associated frequencies are counted. The result is  a set of vectors.  Then, sequences are compared pairwise by computing suitable distance/similarity functions between each pair of vectors. The interested reader can find in \cite{Zielezinski2019} a list of the ones that have been the object of a  recent benchmarking study.  One of the most surprising findings  of that study is that  those simple AF functions are among the best performing and most versatile  in terms of application domain. We have chosen the best performing ones according to that benchmarking, representatives of all types of AF functions described in  \cite{luczak2017survey} and that can be broadly used in biological studies, e.g., metagenomics \cite{Benoit16}. The complete list of the selected AF functions is in Section 2.1 of the \SM, together with their definitions.

An AF function in the second class is based on  the notion of {\em match} between two sequences. This latter is  usually encoded via a binary vector, where the one entries indicate  the positions where two subsequences of  the two sequences must be identical. Zero entries may not  matter. Hence, the distance between two sequences is mainly estimated according to the length of their substring matches. The interested reader can find examples of those methods in \cite{leimeister2014fast,leimeister2014kmacs,morgenstern2015estimating}. For our study, we have chosen {\em spaced word methods} \cite{Leimester17}. In particular, the {\bf FSMW} distance, since it has emerged as the most competitive in this class of AF functions \cite{Leimester17,Zielezinski2019}, details regarding its definition are in Section 2.2 of the \SM.

\subsection{A Spark Platform for Fast Computation of AF Functions: the Basic Pipeline} \label{subsec:framework}

In this section, we first  provide a \vir{user level} functional description of the proposed platform. Then we outline, again at a functional level, how the architectural issues influencing scalability have been dealt with.

\subsection{A User-view of the Basic Pipeline as a General and Extensible Spark Programming Paradigm for Implementation of AF Functions}
\label{subsubsec:basicpipeline}

To a user, the basic pipeline appears as a  succession of stages, described next. Assuming that the dataset to be processed is composed of $n$ sequences, the output is an $n \times n$ matrix, in which entry $(i,j)$ corresponds to the value of the chosen AF function on sequences $i$ and $j$.   It is also worth pointing out that since  the input sequences are partitioned over different computing nodes, two steps are required to collect a global statistic. First, the desired statistic is partially evaluated on each node holding a part of a given sequence. Second, all partial statistics are aggregated to derive the global statistic.

\begin{itemize}
	\item \textit{Stage 1: Collection of Partial Statistics.}  In this stage, the statistic that needs to be collected, e.g. \kmers, is  extracted from each of the input sequences and provided as output.
	This is transparently done in a distributed way, so that each computing node  extracts  the partial statistic from the parts of the input sequences it stores. We anticipate that Stage 3 takes care of aggregating the different partial statistics extracted from a same sequence.  	User code can be provided to support more statistics, in addition to those already included in the platform.

	\item \textit{Stage 2: Feature based Statistics Filtering.} The user implementing the AF algorithm may require the exclusion of a selected subset of features from the statistics it is computing, e.g., specific $k$-mers such as those containing the \vir{N} character. To this end, this stage acts as a filter to exclude from the output of the  previous stage the selected features, according to conditions specified by the user. The filtering occurs at this point, so as to (possibly) alleviate the workload of the following stages.

	\item \textit{Stage 3: Statistics Aggregation.} All partial statistics extracted by different computing nodes during Stage 1 (and possibly Stage 2),  but originating from the same input sequence are automatically and transparently gathered on a same node and aggregated. For instance, statistics about a particular \kmer and extracted from different parts of the same sequence are summed to obtain the overall \kmers statistics for that sequence. User code can be provided to support aggregation for statistics, in addition to those already included in the platform.
	 
	\item \textit{Stage 4: Value-based Aggregated Statistics Filtering and Normalization.} Stage 2 filters the features existing in a statistic, while this stage filters according to a user-defined condition targeting the aggregated value assumed by a feature in a statistic. For instance, one would want to exclude low frequency \kmers when collecting \kmers statistics. This stage also performs, if required, data normalization. Indeed, as well argued in \cite{luczak2017survey}, it is advisable to take the statistic of each sequence, e.g., \kmers counts, and transform it so that all the statistics refer to the same scale. Details are in  \cite{Giancarlo18,luczak2017survey}). 
	
		\item \textit{Stage 5: AF matrix computation.} For each pair of different input sequences, their final aggregated statistics are sent by the platform  to the same node. 
 The AF function that has been chosen,  from the ones available,  is evaluated on  each pair of sequences and the AF matrix is filled accordingly.  In addition to the ones available, additional  functions can be supported by providing user code. 
\end{itemize}

The output AF matrix  is encoded as a distributed data structure, whose content can be saved on file or used as input for further analysis. 

Each of the aforementioned stages is modeled as one or more Spark distributed transformations. A general and extensible library of built-in basic functions implementing them is described in details in Section 3.1 of the \SM. The user interested in supporting a new statistic and/or implementing a variant of these functions can provide her code, as described in Section 3.2 of the \SM.

\medskip

\subsection{Architectural Engineering: Tuning the Pipeline as a Function of the Workload}
\label{subsec:tuning}

We briefly highlight the different data partitioning strategies  supported by FADE, designed with the aim of tuning the  basic pipeline as a function of the input workload so as to allow for an efficient and scalable execution. Additional details  regarding them  are available in Section 4 of the \SM, while their  comparative experimental evaluation   is reported  in Section \ref{sec:intuition3}.

\begin{itemize}

	\item \textit{Strategy 1: Total aggregation.}  This strategy allows for very good execution times when extracting and processing statistics having an overall small size. This is possible because all statistics (either partial or aggregated) extracted during the pipeline are maintained and processed on a single node of the distributed system. The same occurs to the partial AF function  evaluated on each statistic. On a one side, this implies that no distributed computation occurs, apart from that of Stage 1. On the other side, this strategy allows to avoid the data transmission overhead required to transfer data to the nodes of the distributed system prior to their processing.

	\item \textit{Strategy 2: No aggregation.} This strategy allows for a very good scalability when extracting and processing statistics from very large input data. This is possible because every single statistic (either partial or aggregated) extracted during the pipeline is managed as a stand-alone data object. The same occurs to the partial AF function  evaluated on each statistic. The only aggregation occurs at the end of the pipeline when, for each pair of distinct  sequences, partial AF function values  are combined to return the overall value of the function. This ensures for a very fine scalability and load balancing as Spark tends to scatter these data objects uniformly at random on the different nodes of the distributed system.  This holds because the amount of memory required to process single data objects is, typically, much smaller than the one required for processing collections of data objects. 

	\item \textit{Strategy 3: Partial aggregation.} This strategy allows for a good trade-off between efficiency and scalability when extracting and processing statistics from  large input data. This is possible because all statistics (either partial or aggregated) extracted during the pipeline are partitioned into bins. The same occurs to the partial AF function evaluated on each statistic. Consequently,  each node processes  a smaller number of data records batches (i.e., the content of each bin) rather than a (potentially) much larger number of single data records. This has  a positive effect both on the processing and the communication times. 
	\end{itemize}

\subsection{Reliability and Robustness of AF Functions}
\label{subsec:reliability}

\subsubsection{Reliability of an AF Function  via a Hypothesis Test Monte Carlo Simulation}\label{sec:statsign}
Intuition suggests that the larger the number of entries in the AF matrix not due to \vir{chance}, the more indicative of biological relevance the outcome, e.g. phylogenetic tree, of that function use is expected to be. However, experience suggests that AF functions may have a \vir{behavior} that depends on the dataset being processed.  Therefore, it is uncontroversially desirable to use  functions that are consistently significant.

 We formalize such an intuition by requiring that a  consistent AF function must provide  a high percentage of statistically significant entries in its  corresponding AF matrix, with very little dependence  on  the input dataset. In regard to those tests,  we resort to a Monte Carlo Simulation method, in which we assume as the Null Hypothesis $H_0$ that two biological sequences are as similar as two random ones. An entry $(i,j)$ of an AF matrix is statistically significant  when the Null Hypothesis involving the two sequences associated to that entry is rejected with a given significance level.  The entire simulation consists of three steps, described next. The Spark algorithms implementing it are briefly described in Section 5 of the \SM. 
In what follows, we consider only the case of similarities, since the case of distances is analogous.

\paragraph{Step 1:Synthetic Datasets Generation via Bootstrapping.}\label{sec:boot}
The first step is to devise a procedure that generates synthetic datasets, that are meant to represent \vir{random data}. Such a task can be accomplished by choosing a \emph{Null Model}, e.g., an Information Source emitting symbols uniformly and at random. However, in our case, it seems more appropriate to resort to \emph{bootstrapping}  \cite{efron}, i.e., to generate the synthetic datasets from real ones, since it is desirable to preserve the biological origin of the input dataset also in the synthetic ones.  To this end, we proceed as follows.

Let $S$ be the input dataset and let $q$ be a parameter. All $q$-mers of sequences in $S$ are extracted and placed in a bin $B$. Then, in order to obtain a synthetic dataset  $\hat S$, we extract uniformly and at random $q$-mers from $B$ in order to form new sequences to be included in $\hat S$. This latter has the same number of sequences as in $S$ and each sequence in $\hat S$ corresponds to only one in $S$ in terms of length. 

It is to be noted that the parameter $q$ allows us to generate synthetic datasets along a wide spectrum of subsequence  statistics  present in $S$, e.g., $q=1$ corresponds to the case in which the synthetic dataset is generated according to the empirical probability distribution  of symbols in $S$.  

\paragraph{Step 2: Significance Test via a Monte Carlo Simulation for Two Sequences.}\label{sec:sign-pair}
The next step consists of the following simulation, adapted from \cite{Giancarlo08}. It applies to sets of two sequences. 
$C$ denotes the AF function to which the procedure is applied. The Null Hypothesis $H_0$ is that two input sequences are as similar as two random ones, when similarity is assessed via $S$. We want to reject the Null Hypothesis with  confidence level $\alpha$, with the performance of $\ell$ Monte Carlo Simulations.

    \centerline{\bf Procedure MECCA$(\ell,  C , S, \alpha)$}

\begin{itemize}
	
	\item [{\bf (1)}] For $ 1 \leq i \leq \ell$, compute a new set  of two sequences $\hat S_i$ according to the  procedure outlined in Step  1. Compute the similarity between the two sequences  in ${\hat S}_i$ via $C$. Let $T_i$ be its value. 
	
	\item [{\bf (2)}] For $1 \leq i \leq \ell$,  sort the $T_i$ values in non-decreasing order and let $SL$ be the corresponding list.
	
	\item [{\bf (3)}] Let $T$ denote the value of  $C$ computed on $S$.
	Let $j$ be the maximal index such that $SL[j] < T$.   Let 
	$\delta=(j/\ell)$.  The $p$-value is then $1-\delta$ and, and letting $\alpha$ be the desired significance level, the hypothesis that the two sequences in $S$ are as similar as two randomly chosen ones 
	is rejected with that significance level if $1-\delta \leq \alpha$. 
	
\end{itemize}

\paragraph{Step 3: AF Matrix Significance via Bonferroni Correction.}
Consider now a set $S$ consisting of $n$ sequences, labeled from $1$ to $n$. Let $F$ be the  $n\times n$ AF matrix for $S$ computed via $C$. In order to assess how statistically significant is  $F$, with family-wise significance level $\alpha$, we can resort to the pairwise application of the simulation procedure outlined in Step 2. Since we are performing $m=\frac{n(n-1)}{2}$ hypothesis tests, we have to correct for rejecting $H_0$ simply by chance. 
Since those tests may not be assumed to be independent, we use the well known Bonferroni correction. That is, for each test performed for each entry of $F$, $H_0$ is rejected with significance level $\alpha/m$. Then,  we reject the Null Hypothesis that   the matrix is no better than one obtained on a set of random sequences, if all $m$ entries pass the test. However, even if the full matrix does not pass the test, such a procedure outlines entries that are statistically significant in terms of similarity values of the corresponding sequences.

\subsubsection{Robustness of an AF Function via a Matrix Perturbation Method}\label{robmethod}

In addition to the reliability of an AF function, it is also important to assess how robust is the function with respect to \vir{noise}, i.e., the number of entries in the AF matrix that are not statistically significant. Informally, we refer to this as the \emph{operational range} of an AF function. In order to empirically estimate this latter, we informally proceed as follows: the larger is the amount of noise injected in a AF matrix returned by a given function, the worse should be the biological relevance of its outcome (e.g., the phylogenetic tree). Then, to measure the operational range of a function, we start from the AF matrix returned by that function on a given dataset and assume as a reference its performance score. Finally, we study its performance variation, assuming it will decrease while increasing the amount of noisy entries. 
More precisely, we assume that the dataset under consideration has a gold standard solution. For this study, it is a phylogenetic tree associated to the species in the dataset. We also use a computational method $G$ to build a phylogenetic tree from an AF matrix and a distance/similarity measure $D$ to assess how different is a tree produced by $G$ via an AF matrix with respect to the gold standard. Formal details of the basic  perturbation step follow.

\begin{itemize}
\item[$\bullet$]	Given the AF matrix computed by $C$ on dataset $S$, we select uniformly and at random  a given percentage of  entries and substitute each value with a \vir{noisy} one. 
	For histogram-based functions, such a value is taken randomly  among the ones appearing in the AF matrices  generated via Monte Carlo simulation. 
	As for  FSWM, since the AF matrices it returns when evaluated on the synthetic datasets are likely to contain null values (see discussion about filtering in Section \ref{subsubsec:intuition1analysis}), the \vir{noisy} value is obtained by increasing the original one by  a value chosen  uniformly and at random. For both types of AF functions, the new matrix so obtained is used to build a tree via $G$, and its distance from the gold standard is computed via $D$. The loss in performance at the given noise level is given by the difference in the $D$ score obtained with the noisy vs the original matrix. 
\end{itemize}

\section{Results and Discussion}\label{subsec:results}
As a  first preliminary step, we selected for our study datasets coming from \cite{Zielezinski2019}, with the criterion that  the  AF functions chosen here would work on them (see again \cite{Zielezinski2019}). They are reported in Section 6  of the \SM. As  a second  preliminary step, we assessed that our AF functions implementations are in line with those used for the benchmarking of the AFproject \cite{AFproject2019} (details are omitted for brevity and available upon request). All our experiments have been executed on the hardware platform described in Section 6 of the \SM. The parameters of the algorithms have been set according to the procedures described in Section 7 of the \SM. Execution times have been measured by collecting the job execution elapsed time returned within the Spark framework.

\subsection{Assessing the Scalability of Methods Supported by FADE and the Effectiveness of The Aggregation Strategies}\label{sec:intuition3}

\subsubsection{Experiments}

When considering a single computing platform where multiple computing units coexist on the same motherboard and communicate, at no cost, by using shared memory, nothing can beat the performance of a native application purposely developed to take advantage of that setting. On the other hand, the number of processing units and memory size on a single motherboard is a natural performance bottleneck. 

Given the above scenario, the use of a distributed framework is 
to overcome such a bottleneck, by scaling  performance as a function of the number of computing units far over the ones on a stand-alone computing platform. Therefore, it is important to assess the efficiency of methods supported by our software pipeline, when compared to analogous state-of-art shared memory alignment-free tools, while assessing  the scalability of our framework. For such an analysis, we have chosen two reference software systems that provide shared memory parallel software for their evaluation: FSWM \cite{leimeister2014fast} and Mash \cite{Ondov2016}.

\begin{figure}
    \centering
    \includegraphics[width=0.85\linewidth]{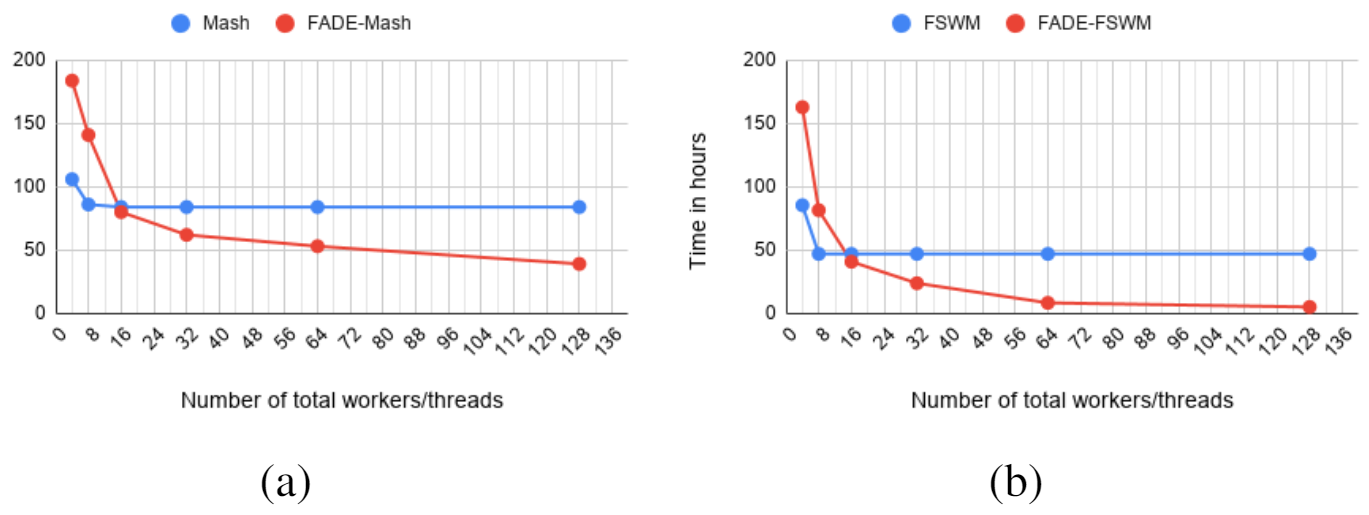}
    \caption{Execution time, on the ordinate axis, required by both versions of  (a) Mash  and (b) FSWM  on the  Plants (assembled), using an increasing number of workers/threads. The Partial Aggregation strategy has been used, since naturally suited to  the algorithmic methods supporting Mash and FSWM, respectively.}
	\label{fig:fade-mash-fswm}
\end{figure}

We measure the execution time of FSWM and Mash, as well as  that of their implementations supported by our framework, i.e., FADE-FSWM and FADE-Mash, on the Plants (assembled) dataset  and while increasing the level of parallelism, as explained shortly. Mash and FSWM have been executed on a single machine equipped with $8$ computing cores. The FADE  versions  have been executed on a distributed framework equipped  with 24 worker nodes, for a total of $128$ computing cores. Each of those machines is identical to the one on which the shared memory algorithms have been executed. The result of this experiment is visible in Figure \ref{fig:fade-mash-fswm}, where we report the execution time of these applications as a function of the number of concurrent threads, for shared memory applications, or of Spark workers, for FADE. Each thread/worker executes on a single core.

In order to assess the effectiveness of the aggregation strategies supported by FADE, we executed a significance test for all of the histogram-based functions, with different data sizes. The results are reported in Table 
\ref{tab:tot_aggr}.

\subsubsection{Results}

The results of the first experiment show that the shared memory versions of Mash and FSWM are faster than the implementations supported by FADE,   when using a very small number of threads compared to workers. This is expected. With the increase of the number of those units, the performance gap narrows. As soon as the use of $16$ threads/workers is reached, FADE-Mash and FADE-FSWM  require about the same execution time as that of  their shared memory counterparts. However, those latter stop scaling, due to the bottleneck mentioned earlier,  while the FADE methods  continue to scale as  more threads/workers  are added.

The results of the second experiment clearly show that the Partial Aggregation strategy is the most flexible of the three, since it guarantees equal or better performance with respect to the other two, independently of the workload. 
\begin{table}[!ht]
		\centering
		\begin{tabular}{|r|r|r|r|r|}
			\hline
			Dataset & No Aggregation & Partial Aggregation & Total Aggregation\\
			\hline
			Small & 0.9 & 0.9 & 1.1 \\
			Large & 2.73 & 2.82 & NA \\
			Very large & 48.3 & 42.0 & NA \\
			\hline
		\end{tabular}
		\caption{Execution time, in minutes, required by our framework to execute one instance of the AF significance test, for all of the histogram-based functions,  on three reference datasets with different aggregation strategies. The NA value indicate that the test took too long to complete and was stopped.  Small = Mithocondria (assembled). Large = Shigella (assembled). Very large = Plants (assembled).}	
		\label{tab:tot_aggr}
\end{table}

\begin{figure}
	\centering
	\includegraphics[width=400pt]{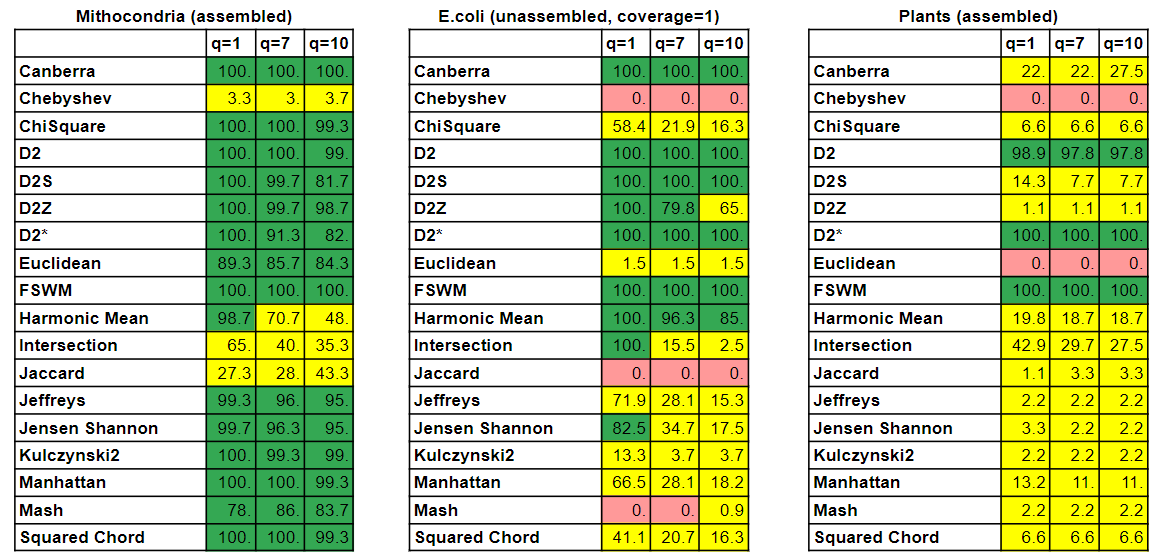}
	\caption{Summary of the Hypothesis Test results for the different AF functions considered in this research  when executed on three different datasets with $q={1,7,10}$ and with significance level set to $1\%$. 
	Datasets have been chosen as described in the Main Text.  \label{fig:sigtestall}}
\end{figure}

\subsection{Reliability and Robustness of AF Functions: An Application of FADE to  Data/Compute Intensive Analysis}
\label{subsec:reliabilityexp}

\subsubsection{Reliability of AF Functions}\label{subsec:sigtest}

\paragraph{The Experiment.}

\medskip
For each of the benchmark datasets included in this study, we execute the AF statistical significance test described in Section \ref{sec:statsign}. 
Specifically, for each  matrix obtained via a dataset,  we generate additional ones via bootstrapping. Then, we reject the null hypothesis family-wise with $p$-value $\leq 1\%$, applying Bonferroni correction to all of its  $m$ entries. The number $l$ of simulations for each test has been chosen according to the size of the dataset being processed, so as to guarantee the execution of the experiments in a reasonable amount of time. An outline of these parameters, including the choice of $k$ and the values of $q$,  is available in Section  7 of the \SM.

A summary of the results is shown in Figure 10 of the \SM. For each dataset, each AF function and each considered  null model,  the percentage of entries passing the test is reported. This value is drawn in green, if at least  75\% of the entries passes the test, in red, if no entry passes the test,  and in yellow, in the remaining case. 
Figure \ref{fig:sigtestall} reports a representative synopsis of those results:  a dataset where most AF functions  perform well, one in which many  perform  poorly and one in which most perform poorly. 
    
\paragraph{Results and Insights}
\label{subsubsec:intuition1analysis}
\begin{itemize}
	\item{ } {\em A novel class of AF functions:consistently significant}.  With reference to Figure \ref{fig:sigtestall} and Figure 10 of the \SM, it is evident that there are AF functions returning matrices passing the family-wise  significance test either fully or with a high percentage of entries,  for all of the benchmark datasets. We denote them as \textit{consistently significant}, since they behave consistently well, independently of the  dataset and, as already stated in the Introduction, it is well known that statistical relevance is a good indication of biological relevance \cite{DidFr02,  Giancarlo2008,speed96}. They are:  D2, D2*, FSWM.  The remaining ones either perform inconsistently or poorly (Chebyshev).

\item{} {\em Filtering can be useful.}  While the statistical guarantees of AF functions in the D2 family are well known and have been identified via deep investigations \cite{kai2013}, we find a novel fact regarding FSWM: it is quite good in delivering matrices that pass the statistical significance test. This can be attributed to the filtering mechanism present in the algorithm and that was designed to \vir{flush out} the parts of the statistics it is collecting and that are considered \vir{weak}. Such a filtering is able to detect  the \vir{low relatedness} of the synthetic genomes ensuring that the original genomes \vir{win} the test. 




\end{itemize}

\subsubsection{Robustness  of AF functions: the case of consistently significant functions}
\label{subsec:sensitivity}

\ignore{
\begin{table}
\centering
\begin{tabular}{|c|c|c|c|c|c|c|c|c|c|c|c|c|c|c|c|}
\hline
& \multicolumn{5}{|c|}{ q=1 } & \multicolumn{5}{|c|}{ q=7 } & \multicolumn{5}{|c|}{ q=10 } \\
\hline
& \textbf{D2} & \textbf{D2S} & \textbf{D2Z} & \textbf{D2*} & \textbf{FSWM} & \textbf{D2} & \textbf{D2S} & \textbf{DZ} & \textbf{D2*} & \textbf{FSWM} & \textbf{D2} & \textbf{D2S} & \textbf{DZ} & \textbf{D2*} & \textbf{FSWM}\\
\hline
\textbf{Yersinia}     & 4\% & 4\% & 6\% & 6\% & 4\% & 4\% & 4\% & 6\% & 6\% & 4\% & 4\% & 4\% & 6\% & 6\% & 4\%\\
\textbf{Mitochondria} & 2\% & 4\% & 2\% & 2\% & 2\% & 2\% & - & 2\% & - & 2\% & 2\% & - & 4\% & - & 2\%\\
\textbf{Shigella}     & 2\% & 2\% & 2\% & 2\% & 2\% & 2\% & 2\% & 2\% & 2\% & 2\% & 2\% & 2\% & 2\% & 2\% & 2\%\\
\hline
\end{tabular}
\caption{Minimum percentage of corrupted entries observed, for each dataset, each value of $q$ and each AF function, to make the score of the resulting AF matrix worse than the reference score by, at least, a 10\%. The dash symbol (-) indicates that the corresponding AF function is not statistically significant on that specific dataset and value of $q$, i.e. has a value $<$ 75\% in significance test.}
\label{tab:fine_grained_analysis}
\end{table}

}

\paragraph{The Experiment.}
\label{subsubsec:exp2}
We concentrate only on AF functions that have been classified as consistently significant in the previous section, since they are the most likely to be useful on a day-to-day basis.  
For each of them, this experiment is conducted by injecting an increasing percentage of noisy entries into an AF matrix, following the method  outlined in Section \ref{robmethod}.  As a computational method $G$ to build a philogenetic tree from an AF matrix,  we use  both UPGMA  and  NJ. This choice is motivated by the fact that, with pros and cons, those two methods are reference points in the Literature.   Moreover,  as a distance/similarity measure $D$ to assess how different is a tree produced by $G$ via an AF matrix with respect to the gold standard, we use both the Matching Cluster  \cite{matchingmetric} (MCM for short) and the Robinson Fould  \cite{robinson1981comparison} (RF, for short)  metrics. Such a choice is motivated by the fact that they were among the top performing ones in a classic benchmarking study \cite{Kuhner}, in particular the second. As for RF, it is quite sensitive to small changes in tree topology, i.e., it \vir{saturates} rapidly, a fact that has been criticized in the past \cite{Penny1982}. Yet, it is a standard in the Literature.  
    
    The results of this experiment are available in Figures 11-16 of the \SM, where we plot the \vir{distance} from the gold standard as the number of  noisy entries grows. 
    Moreover, to gain more insights, we also report in Tables  \ref{T:variation1}-\ref{T:variation3},  
    the variation in distance only when $10\%$ noisy entries are injected. Moreover, we do not consider changes in the distance from the gold standard that are in the interval +/-$3$, i.e., reasonably close to the case of \vir{no noise}.

\paragraph{Results and Insights}
\begin{itemize}
\item {\em AF matrices are sensitive to the injection of noisy entries.} Figures 11-16 of the \SM show a trend of deterioration in classification ability as the percentage of noisy entries increases. Such a trend is more pronounced when we use the RF  metric to assess how close to the gold standard is a philogenetic tree computed from a noisy matrix  with respect to the one that uses an uncorrupted AF matrix. This is possibly due to the \vir{saturation} effect, but MCM  largely confirms those experiments.

The results in Tables  \ref{T:variation1}-\ref{T:variation3} give  more compelling evidence of the sensitivity to noise of the AF functions under analysis. In particular, FSWM is quite sensitive, with both NJ and UPGMA. As for the other two AF functions, they display such a sensitivity mostly when used with NJ. Or, better, UPGMA seems to be able to tolerate a small amount of noise in the input data. We believe that such an aspect of UPGMA is novel. It is to be noted that the entries with negative values, i.e., an improvement in classification with the addition of random entries,  is justified by the fact that the original classification was so poor  that even noise could do no harm.

In conclusion, the indication we receive from the above is to use D2, D2* and FSWM in conjunction with UPGMA and with a high number of statistically significant entries in the AF matrix.   

\end{itemize}

\begin{table}[!ht]
    \centering
    \includegraphics[scale=0.4]{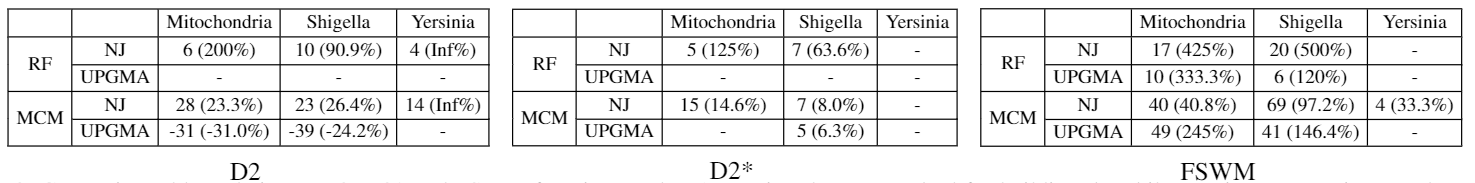}
    \caption{Corruption tables relative to D2, D2* and FSWM functions and q=1, varying dataset, method for building the philogenetic tree, metric to evaluate the distance between the reference and obtained tree. In each entry, it is reported the difference between the philogenetic distance evaluated on the distance matrix with 10\% of corrupted entries and the original distance matrix (between the brackets is reported the percentage difference, indicating with 'Inf' when the tree obtained from the original distance matrix is equal to the reference tree, i.e. the distance between the trees is 0). Dashed entries represent cases where the introduction of noise caused  only a small change in distance, as specified in the Main Text. }\label{T:variation1}
\end{table}

\begin{table}[!ht]
    \centering
    \includegraphics[scale=0.4]{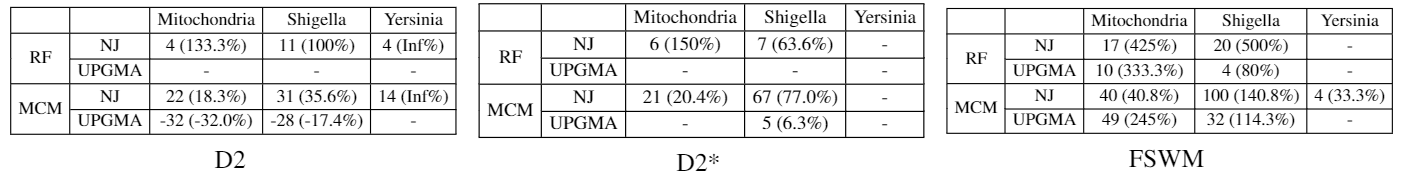}
    \caption{Corruption tables relative to q=7, The Legend is as in Table \ref{T:variation1}}\label{T:variation2}
\end{table}

\begin{table}[!ht]
    \centering
    \includegraphics[scale=0.4]{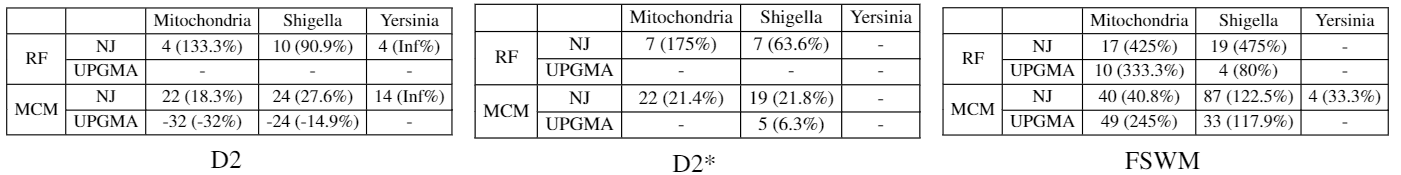}
    \caption{Corruption tables relative to q=10, The Legend is as in Table \ref{T:variation1}}\label{T:variation3}
\end{table}

\section{Conclusion}
We have provided the first Spark platform,  supporting  AF functions evaluation,  that is extensible and that guarantees good scalability with respect to the workload and computational resources available. This is witnessed by the very good performance, both in terms of computing time and scalability, of FADE-Mash and FADE-FSWM when compared to their state-of-the-art shared memory counterparts. This is particularly relevant, since the shared memory  version of Mash has a leadership position due to its speed.

In order to demonstrate the ability of FADE to support large data and compute intesive tasks, we have conducted a novel analysis of AF functions. Thanks to it, we gain insights into which functions are best suited for day-to-day AF genomic analysis: D2, D2* and FSWM. We also account for the  \vir{operational range} of those latter measures, i.e., circumstances in which a result due to those methods can be trusted. Such an \vir{operational range} translates into the requirement to use the AF matrices corresponding to those methods when the vast majority of their entries is statistically significant. Moreover, the use of UPGMA is recommended, since it is less sensitive to \vir{noise} in the input data, compared to NJ.

\section*{Acknowledgements}
R.G. is grateful to Prof. Chiara Romualdi for helpful discussions. All authors would like to thank the Department of Statistical Sciences of University of Rome - La Sapienza for computing time on the TeraStat cluster and for other computing resources, and the GARR Consortium for having made available a cutting edge OpenStack Virtual Datacenter for this research.

\section*{Funding}
G.C., R.G. and U.F.P. are partially supported by GNCS Project 2019 \vir{Innovative methods for the solution of medical and biological big data}. R.G. is additionally supported by  MIUR-PRIN project \vir{Multicriteria Data Structures and Algorithms: from compressed to learned indexes, and beyond} grant n. 2017WR7SHH. 
U.F.P. and F.P. are partially supported by Universit\`{a} di Roma - La Sapienza Research Project 2018 \vir{Analisi, sviluppo e sperimentazione di algoritmi praticamente efficienti}.

\bibliographystyle{abbrv}
\bibliography{main}

\end{document}


\maketitle
			
\abstract{
Additional details about the \MM are provided in this document.
}

\maketitle

\section{Spark and Distributed Architectures}\label{sec:apache} 
Apache Spark \cite{spark} is a  framework used mainly to support programs with in-memory computing and acyclic data-flow model to be executed on a distributed computing architecture. In simple and intuitive terms, this latter can be described as a set of computing nodes  that cooperate in order to solve a problem via local computing and by exchange of \vir{messages} \cite{lynch1996distributed}.  On that type of architecture,  Spark can be used for applications that reuse a working set of data across multiple parallel operations (e.g., iterative algorithms) and it allows the combination of streaming and batch processing, as opposed to Hadoop \cite{hadoop}, that can be only used for batch applications.
In addition, Spark is not limited to support only the \MR \cite{dean2008mapreduce} paradigm, although it preserves all the facilities of \MR-based frameworks, and it can have the Hadoop as the underlying  middleware.


It provides high-level APIs which support multiple languages (Scala, its native language, Java and Python) and is made of two fundamental logic blocks: a programming model that creates a task dependency graph, and an optimized runtime system which exploits this graph to deploy code and data and schedule work units on the distributed system  nodes. At the core of the Spark programming model is the \emph{Resilient Distributed Dataset} (RDD) abstraction, a fault-tolerant, distributed data structure that can be created and manipulated using a rich set of operators: programmers start by defining one or more RDDs through \emph{transformations} of data that originally resides on stable storage or other RDDs.




\section{A Selection of Alignment-free Distances and Similarity Functions}\label{sm:sec:funct}

In this section, we provide definitions of the AF functions included in this study.

\ignore{
AF functions that are based on {\em \kmer statistics} can be used to analyze  a set of sequences  as follows. For each sequence in the set, the contiguous subwords of length $k$ therein contained (i.e., \kmers) with their associated frequencies are counted. The result is  a set of vectors.  Then, sequences are compared pairwise by computing suitable AF functions,  between each pair of vectors. The interested reader can find in \cite{Zielezinski2019} a list of the ones that have been recently the object of a benchmarking study.

Regarding  {\em word matches}, an  AF function in this class and  computed on two sequences is based on  the notion of {\em match}. This latter is  usually encoded via a binary vector, where the one entries indicate  the positions where two subsequences of  the two sequences must be identical. Zero entries may not  matter. Hence, the distance between two sequences is estimated according to the length of their substring matches. The interested reader can find examples of those methods in \cite{leimeister2014fast,leimeister2014kmacs,morgenstern2015estimating}. 

Among the many possible existing AF functions, our choices for this study are  as follows.  

\begin{itemize}

\item{\em{\kmer statistics: histogram statistics}} \cite{luczak2017survey}.  Such a choice is motivated by the fact that one of the most surprising findings in the extensive benchmarking study presented  in \cite{Zielezinski2019} is that  those simple AF functions are among the best performing and most versatile  in terms of application domain. We have chosen the best performing ones, representatives of all types of AF functions described in  \cite{luczak2017survey} and that can be broadly used in biological studies, e.g., metagenomics \cite{Benoit16}. The complete list of the selected AF functions is in Section \ref{subsec:hist_stat}, together with their definition. 

\item{\em{Word matches: spaced word methods}} \cite{Leimester17}.  Such a choice is motivated by the fact that in this class of AF functions, they have emerged as the most competitive \cite{Leimester17,Zielezinski2019}.  A spaced word method works as follows. In order to compare a pair of sequences, a binary pattern is introduced, to distinguish among significant positions (match) and not significant positions (don't care). Then, the distance between the two sequences is evaluated by computing a similarity score according to the similarity of their don't care positions and to the number of substitutions required to perfectly align the two sequences in the match positions. More details about the AF function are in Section \ref{subsec:sw}.


\end{itemize}


 
}
 \subsection{Histogram Statistics Selected for this Study}
 \label{subsec:hist_stat}
 In what follows, we adopt both the classification and the notation from \cite{luczak2017survey}. It is to be remarked that  some of the AF functions defined next appear in the Literature with different names or they are easy variants of the functions introduced here.  For instance, FFP adopted in 
 \cite{sims2011whole} is the well known Jensen-Shannon Divergence defined here, while Skmer  \cite{Bafna19} is a fast approximation method for the computation of the well known Jaccard Index defined here. 
 For the interested reader, the original publication where the functions defined here have been introduced can be found in \cite{luczak2017survey} for the less known cases.

 Given a set of  sequences $S = \{s_1, ..., s_n\}$,  a \kmer   histogram $h_{s}$  for a sequence $s$ in the set is defined as follows:
 \begin{equation}
  h_{s} = \langle c(w_1), c(w_2),..., c(w_{|K|}) \rangle
\end{equation}
\noindent where $c(w_i)$ is the number of occurrences of the word $w_i$ (i.e. the $i$-th \kmer) in the sequence $s$ and $K$ is the set of all possible words of length $k$ over the alphabet $\{A, C, G, T\}$.

\subsubsection{The Minkowski Family}
 Given two sequences $s$ and $t$ and their associated statistics $h_s$ and $h_t$, the \emph{Euclidean} distance  is defined as:

\begin{equation}
\label{euclidean}
Euclidean(h_s, h_t) = \sqrt{\sum_{w \in K} (h_s(w) - h_t(w))^2}
\end{equation}

\noindent A widely adopted variant of the \emph{Euclidean} distance is the \emph{Manhattan} distance:

\begin{equation}
Manhattan(h_s, h_t) = \sum_{w \in K} |h_s(w) - h_t(w)|
\end{equation}

\noindent Another member of this family, the \emph{Chebyshev} distance, is based on the idea of applying the $p$-th root on the \emph{Manhattan} distance, with $p \to \infty$, and it is defined as follows:

\begin{equation}
Chebyshev(h_s, h_t) = \max_{w \in K} |h_s(w) - h_t(w)|
\end{equation}

\subsubsection{The Match/mismatch  Family}
Let $h^*_s$ and $h^*_t$ be the sets of \kmers with non-zero entries in $h_s$ and  $h_t$, respectively
The \emph{Jaccard} index measures the similarity between those two sets as follows:

\begin{equation}
\label{jaccard}
Jaccard(h_s, h_t) =  \frac{(h^*_s \cap h^*_t)}{(h^*_s \cup h^*_t)}
\end{equation}

It is to be noted that here we use the \vir{classic} definition of Jaccard index rather than the one given in  \cite{luczak2017survey}. Moreover, the computation of the Jaccard index has been the object of much attention in the Information Retrieval community and recently in Bioinformatics. Therefore, we also consider the MinHash approximation algorithm presented in \cite{Ondov2016}. The basic idea is  to reduce the input sequences to sets of integers of small cardinality, each referred to as a \emph{sketch}. To create a single sketch of cardinality  $s$ (i.e., the \emph{sketch size}), each \kmer  in a sequence is hashed. Such a process assigns \kmers to integers pseudorandomly. The $s$ smallest values are retained. Then, in order to compute the Jaccard index of two sequences, the intersection of their sketches is computed. Indeed, the percentage of shared elements in the sketches can be shown to be an approximation of the Jaccard index (for details, see \cite{Ondov2016}).

\subsubsection{The $\chi^2$ Distance} 
It is defined as:
\begin{equation}
\chi^2(h_s, h_t) = \sum_{w \in K} \frac{(h_s(w) - h_t(w))^2}{(h_s(w) + h_t(w))}
\end{equation}

\subsubsection{The Canberra Distance}
It is a mix between Manhattan and $\chi^2$ distances:

\begin{equation}
Canberra(h_s, h_t) = \sum_{w \in K} \frac{\lvert h_s(w) - h_t(w) \rvert} {(h_s(w) + h_t(w))}
\end{equation}

\subsubsection{The $D_2$ Statistics}

It  expresses the similarity of two sequences in a very natural way as a sort of inner product between two histograms as shown in equation (\ref{d2}).

\begin{equation}
\label{d2}
D_2(h_s, h_t) = \sum_{w \in K} {h_s(w) h_t(w)}
\end{equation}

However, extensive studies of this function suggest that a standardized version of it is more useful in the AF sequence analysis setting.
Such a variant is denoted $D_2^Z$. It is as $D_2$, except that the histograms have been standardized via the  well known z-score transformation. Details can be found in \cite{luczak2017survey}. 

We also consider two other members of the $D_2$ family, denoted $D_2^S$ and $D_2^*$ and described in \cite{song2013new}.

$D_2^S$ is based on the finding \cite{shepp1962} that if two independent random variables $X$ and $Y$ are normally distributed with mean zero, also $\frac{XY}{\sqrt{X^2 + Y^2}}$ is normally distributed. We normalize $h_s(w)$ and $h_t(w)$ as follow:

$$\tilde{h}_s(w) = h_s(w) - np_s(w)$$ and $$\tilde{h}_t(w) = h_t(w) - mp_t(w)$$

In which $n$ and $m$ are the number of \kmers, respectively, in $S$ and $T$, while $p_s(w)$ and $p_t(w)$ are the probabilities of the \kmer $w$ under the background model for, respectively, $S$ and $T$.

The $D_2^S$ statistics is defined as follow:

\begin{equation}
D_2^S = \sum_{w \in K} {\frac{\tilde{h}_s(w) \tilde{h}_s(w)}{\sqrt{\tilde{h}_s(w)^2 + \tilde{h}_s(w)^2}}}
\end{equation} 

The $D_2^*$ statistic is based on the idea that, for relatively long \kmers, the number of occurrences can be approximated by a Poisson distribution, thus mean and variance are nearly the same.

The $D_2^*$ statistics is defined as follow:

\begin{equation}
D_2^* = \sum_{w \in K} {\frac{\tilde{h}_s(w) \tilde{h}_s(w)}{\sqrt{np_s(w)mp_t(w)}}}
\end{equation} 

\subsubsection{The Intersection Family}

From this family, we selected two distances. The first is the \emph{Intersection} distance, also known as \emph{Czekanowski}. It   is based on the intersection of the \kmers counts divided by their  union.  It is: 

\begin{equation}
\label{inter}
Intersection(h_s, h_t) = \sum_{w \in K} \frac{2 * min(h_s(w), h_t(w))} {h_s(w) + h_t(w)}
\end{equation}

The  second is the \emph{Kulczynski2}  distance;
\begin{equation}
\label{kulczynski2}
Kulczynski2(h_s, h_t) = A_\mu \sum_{w \in K} min(h_s(w), h_t(w))
\end{equation}

\noindent where $A_\mu$ is equal to  $\frac{4^k(\mu_s-\mu_t)}{2\mu_s \mu_t}$ ($\mu_s$ and $\mu_t$ denote the mean of the histograms $h_s$ and $h_t$, respectively).

\subsubsection{ The Inner Product Family} 

The \emph{Harmonic Mean} distance is:

\begin{equation}
\label{harmonicmean}
Harmonic Mean(h_s, h_t) = 2  \sum_{w \in K} \frac{h_s(w) h_t(w)}{h_s(w) + h_t(w)}
\end{equation}

As opposed to the  Euclidean distance that  computes  the square root over the summation value,  the \emph{Squared Chord} computes the square root over each value of the histograms independently:. 

\begin{equation}
\label{squaredchord}
Squared Chord(h_s, h_t) = \sum_{w \in K} \left(\sqrt{h_s(w)} - \sqrt{h_t(w)}\right)^2
\end{equation}

This can be simplified as
\begin{equation}
 = \sum_{w \in K} {h_s(w) + h_t(w) - 2 \sqrt{h_s(w) h_t(w)}}
\end{equation}

\subsubsection{The Divergence Family}

This family uses probabilities to compare two sequences measuring the distance in the log-probability space. From such a family,  we selected the \emph{Jeffrey} and the \emph{Jensen-Shannon} distances.
The former is defined as:
\begin{equation}
\label{jeffrey}
Jeffrey(h_s, h_t) = \sum_{w \in K} \left(p_s(w) - p_t(w)\right) \ln{ \frac{p_s(w)}{p_t(w)}}
\end{equation}

while the second (JSD for short) is defined in (\ref{jsd}):
\begin{equation}
\label{jsd}
    \begin{split}
    JSD(h_s, h_t) = \frac{1}{2}\sum_{w \in K} p_s(w) \log_2 \left(\frac{p_s(w)}{\frac{1}{2}\left( p_s(w) + p_t(w)\right)}\right)\\
    + \frac{1}{2}\sum_{w \in K} p_t(w) \log_2 \left(\frac{p_t(w)}{\frac{1}{2}\left( p_s(w) + p_t(w)\right)}\right)
    \end{split}
\end{equation}

\noindent where $p_s(w)$ is the empirical probability of \kmer $w$ over all the strings of length $k$ from the alphabet $\{A,C,G,T\}$ in the input sequence $s$ and therefore $p_s(w) = \frac{h_s(w)}{4^k}$.

\subsection{Spaced Words Methods Selected for this Study}
\label{subsec:sw}

\subsubsection{The FSWM Distance}
\label{subsubsec:fswm}

\textit{Filtered Spaced Word Matches (FSWM)} is a method introduced in \cite{Leimester17}.

We need to define a spaced word match between two sequences $s$ and $t$. Given a length $l$ binary pattern $P$ of \textit{match} and \textit{don't care} positions, there is a \sw matching between  $s$, $t$, in positions $i_1$ and $i_2$, respectively,  according to the  pattern $P$, if for each match position $m$ in $P$, it is true that $s[i_1+m]$ = $t[i_2+m]$. This match is also weighted according to the Chiaromonte's substitution matrix \cite{Chiaromonte01}, applied on the \textit{don't care} positions.

All such matches are collected and the ones with a score lower than a given threshold are discarded. Finally, the distance between $s$ and $t$ is computed  by applying \textit{Jukes-Cantor} correction based on the computed  spaced word matches:

\begin{equation}
FSWM(s, t) = -\frac{3}{4} \log \left(1 - \frac{4}{3}\frac{mm_{s,t}}{\delta_{s,t}} \right), 
\end{equation}

\noindent where $mm_{s,t}$ is the number of \textit{don't care} characters that don't match in the collection of spaced word matches between $s$ and $t$, while $\delta_{s,t}$ is the total number of \textit{don't care} characters in that collection.

\section{A General and Extensible Spark Programming Paradigm for Implementation of AF Functions: Details}
\label{userview}

\subsection{User Programming: Software Library and Examples}\label{library}

\subsubsection{The Primitives Library}
\begin{figure}
	\centering
	\caption{FADE class diagram.}
	\label{fig:architecture}
	\includegraphics[angle=-90, scale=0.28]{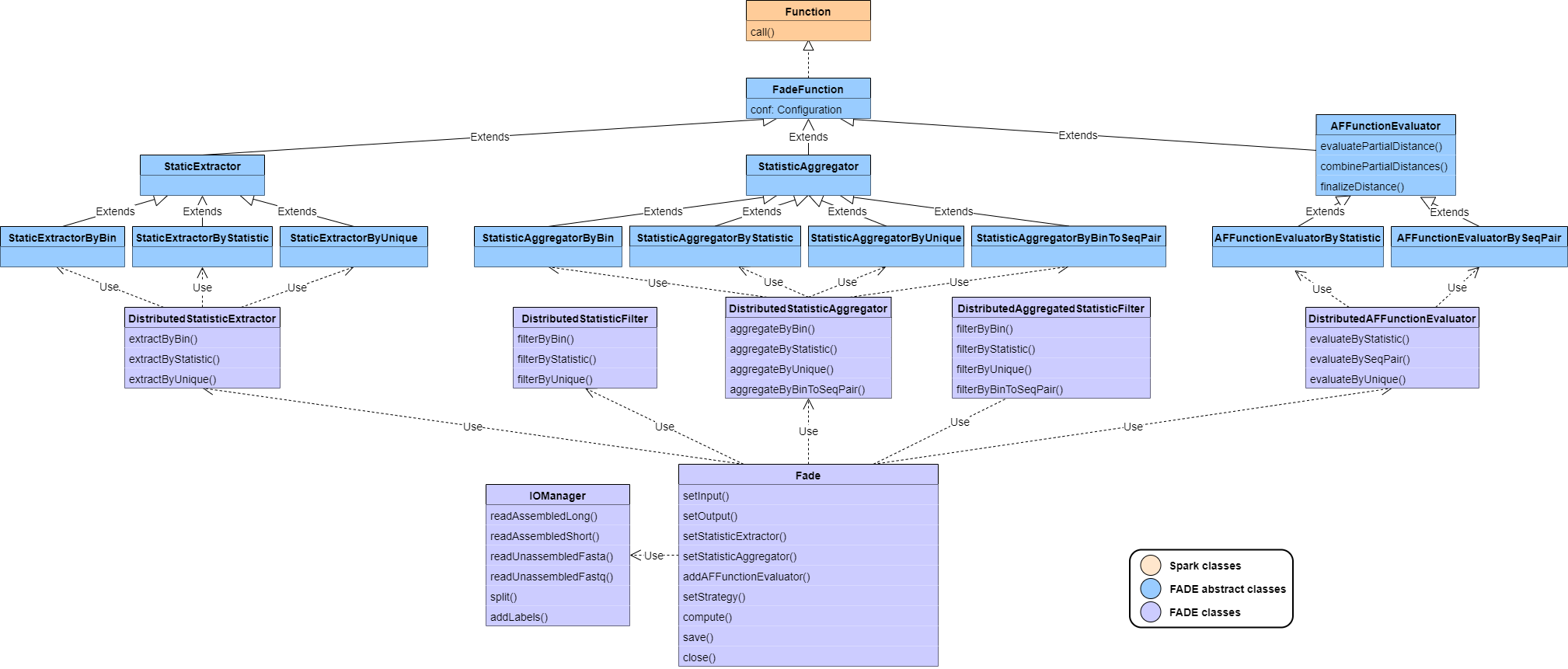}
\end{figure}

This library consists of ready-to-use classes and modules for the computation of AF functions within our system. 
Each of the functions defined in Section \ref{sm:sec:funct} has its own class, identified by the function name.  Each of them  uses  a set of modules, mentioned next,  that refer to the statistics supporting a given function and that naturally correspond to the stages of the basic pipeline (see Figure 1 of the \MM).  The general library is outlined in Figure \ref{fig:architecture}. The main three modules, i.e., \texttt{StatisticExtractor}, 
\texttt{StatisticAggregator}, \texttt{AFFunctionEvaluator}, correspond to stages 1, 3 and 5 of the basic pipeline and are described first.  They are responsible for the layout of the software that is executed, depending on the aggregation strategy that the user specifies. Such a task is transparent to the user and it is described in detail in Section \ref{sec:AED}. In turn, they specialize with respect to the statistics that is relevant for the AF function to be executed, as we outline next.

\begin{figure}
	\centering
	\caption{Kmer specialization class diagram.}
	\label{fig:architecture-kmer}
	\includegraphics[angle=-90, scale=0.5]{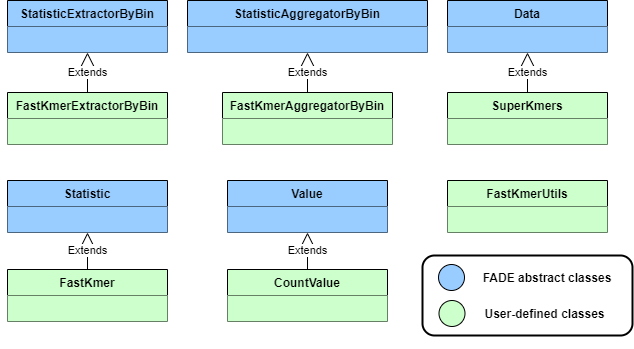}
\end{figure}

\begin{figure}
	\centering
	\caption{Spacedword specialization class diagram.}
	\label{fig:architecture-sw}
	\includegraphics[angle=-90, scale=0.5]{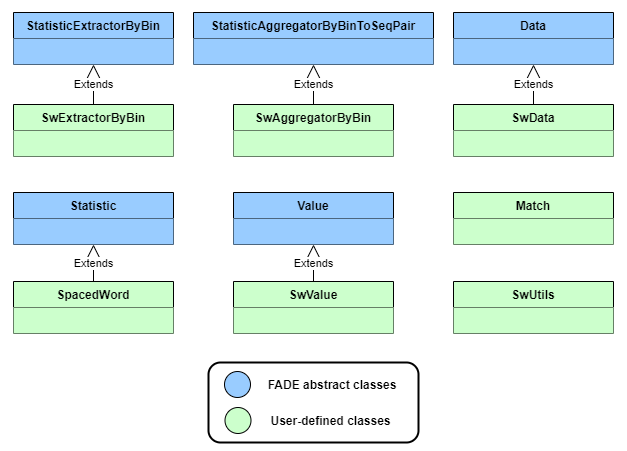}
\end{figure}

\begin{figure}
	\centering
	\caption{Mash specialization class diagram.}
	\label{fig:architecture-mash}
	\includegraphics[angle=-90, scale=0.5]{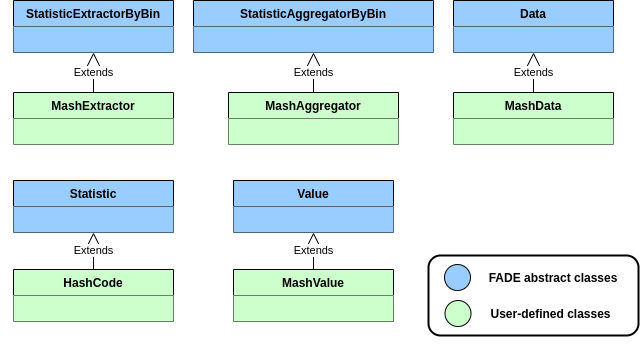}
\end{figure}

\begin{figure}
	\centering
	\caption{AF function specialization class diagram.}
	\label{fig:architecture-dist}
	\includegraphics[angle=-90, scale=0.4]{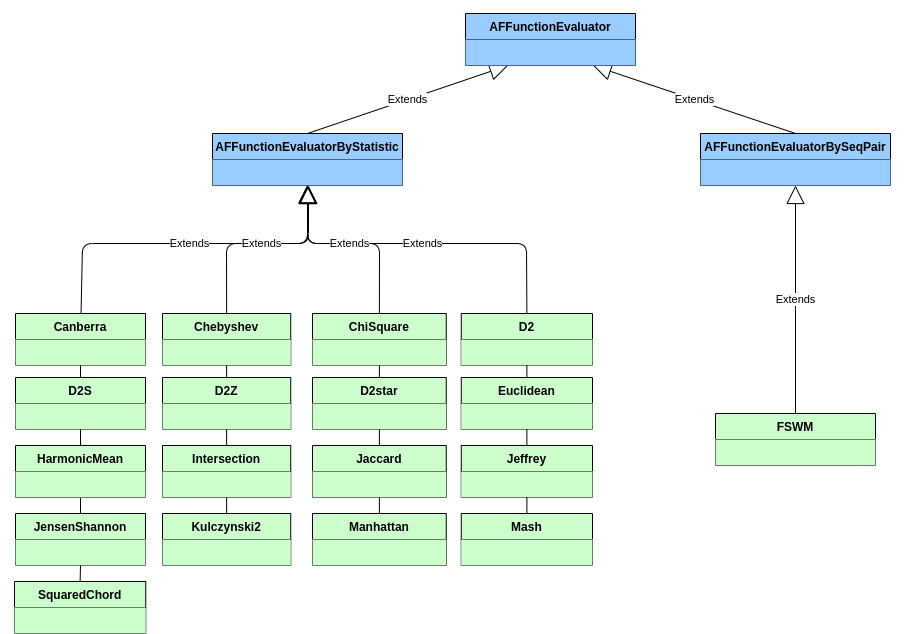}
\end{figure}

\begin{itemize}
    \item  {\em \kmers Histograms}. For this type of functions, the library highlighted in  Figure \ref{fig:architecture} is specialized as shown in Figure  \ref{fig:architecture-kmer}. The modules \texttt{FastKmerExtractorByBin} and \texttt{FastKmerAggregatorByBin} are responsible for \kmer collection and aggregation in accordance with the chosen aggregation strategy. The basic tool to perform this task is the FastKmer package \cite{Fastkmer}. In addition, the modules \texttt{MashExtractor} and \texttt{MashAggregator} are responsible for the execution of \kmers Histograms based AF functions using sketches, such as the MinHash function. This particular type of specialization is shown in Figure \ref{fig:architecture-mash}. It is worth pointing out that one can use a different package for this task, provided that it is properly interfaced with the mentioned classes. A list of modules implementing different distance/similarity evaluation functions by specializing the \texttt{AFFunctionEvaluator} module is reported in  Figure  \ref{fig:architecture-dist}.

    \item  \sws: For this type of functions, the library highlighted in  Figure \ref{fig:architecture} is specialized as shown in Figure  \ref{fig:architecture-sw}. The modules \texttt{SwExtractorByBin} and \texttt{SwAggregatorByBin} are responsible for \sw collection and aggregation in accordance with the chosen aggregation strategy. The distance evaluation function is implemented by the module \texttt{FSWM} reported in  Figure  \ref{fig:architecture-dist}.

\end{itemize}



As for stages 2 and 4, they are implemented by two universal modules \texttt{StatisticFilter} and\\ \texttt{AggregatedStatisticFilter}. They work with any of the statistics supported by our framework, as they operate by evaluating the content and/or the value of each statistic against a user-provided boolean condition encoded using a standard regular expression language. 

\subsubsection{Examples: Euclidean and Filtered SpacedWords Matches Distance}
\label{subsubsec:examples}
Assuming that one wants to evaluate the pairwise  Euclidean distance between sequences originally contained as separate files in the directory {\tt data}, Listing \ref{lst:kmer} provides the Java source code required in our framework to perform that task. It is also possible to achieve the same result by writing the following instructions in a properly formatted configuration file and use it to run FADE (see Listing \ref{lst:kmer_conf}). In such a case, no programming skill is required. 

With reference to Listing \ref{lst:kmer},  the user defines a new Java application where: (i)  the configuration parameters required by the \kmers extraction module are provided, as well as 
the location of the input and of the output files (lines 5-15);  (ii) the choice of which  modules to use for the different stages of the pipeline is indicated (lines 19-22);  (iii)  the resulting pipeline is run on the underlying Spark computing cluster and the output is saved on the chosen location (line 26-28). 

Assuming now that one wants to evaluate the pairwise  Filtered Spaced Word Matches distance \cite{Leimester17} between sequences originally contained as separate files in the directory {\tt data}, Listing \ref{lst:sw} provides the Java source code required in our framework to perform that task. As for the previous case, is also possible to achieve the same result by writing the following instructions in a properly formatted configuration file and use it to run FADE (see Listing \ref{lst:sw_conf}). In such a case, no programming skill is required. 

With reference to Listing \ref{lst:sw},  the user defines a new Java application where: (i)  the configuration parameters required by the \sws extraction module are provided, as well as 
the location of the input and of the output files (lines 5-16);  (ii) the choice of which modules to use for the different stages of the pipeline is indicated (lines 20-23);  (iii)  the resulting pipeline is run on the underlying Spark computing cluster and the output is saved on the chosen location (line 27-29). 

\newpage

\begin{minipage}{\linewidth}
	
	\begin{lstlisting}[language=Java, frame = single, caption={Code required to evaluate the \kmer Euclidean distance among all sequences stored in the data directory with k=13.}, label=lst:kmer, numbers=left, stepnumber=1, xleftmargin=15pt, xrightmargin=20pt]
public class KmerTest {
	public static void main(String[] args) {
	    # Step 1: Input, output and parameters definition
	
	    Configuration conf = new Configuration();
	
	    conf.setInt("k", 13);
	    conf.setInt("x", 3);
	    conf.setInt("m", 6);
	    conf.setInt("slices", 2048);
	
	    Fade f = new Fade(conf);
	
	    f.setInput("data/*.fasta");
	    f.setOutput("distances")
	
	    # Step 2: Modules definition
	
	    f.setStrategy(Strategy.PARTIAL_AGGREGATION);
	    f.setStatisticExtractor(FastKmerExtractorByBin.class);
	    f.setStatisticAggregator(FastKmerAggregatorByBin.class);
	    f.addAFFunctionEvaluator(Euclidean.class);
	
	    # Step 3: Pipeline execution
	
	    f.compute();
	    f.save();
	    f.close();
	}
}
	\end{lstlisting}
\end{minipage}

\begin{minipage}{\linewidth}
	\begin{lstlisting}[language=make, frame = single, caption={Configuration file equivalent to the code reported in Listing \ref{lst:kmer}.}, label=lst:kmer_conf, stepnumber=1, xleftmargin=15pt, xrightmargin=20pt]



# Input, output and parameters definition
k=13
x=3
m=6
slices=2048
task=distance

input=data/*.fasta
output=distances

# Modules definition
strategy=partial_aggregation
extractor=fade.kmer.fast.FastKmerExtractorByBin         
aggregator=fade.kmer.fast.FastKmerAggregatorByBin           
evaluator=fade.affunction.Euclidean

\end{lstlisting}
\end{minipage}

\newpage

\begin{minipage}{\linewidth}
	\begin{lstlisting}[frame=single, caption={Code required to evaluate the \sw FSWM distance among all sequences stored in the data directory with pattern=100101000100011001 and threshold=0.}, label=lst:sw, numbers=left, xleftmargin=15pt, xrightmargin=20pt]
public class SpacedWordTest {
	public static void main(String[] args) {
	    # Step 1: Input, output and parameters definition
	
	    Configuration conf = new Configuration();
	
	    String pattern = "100101000100011001";
	    conf.setString("pattern", pattern);
	    conf.setInt("threshold", 0);
	    conf.setInt("k", pattern.length());
	    conf.setInt("slices", 2048);	
	
	    Fade f = new Fade(conf);
	
	    f.setInput("data/*.fasta");
	    f.setOutput("distances");
	
	    # Step 2: Modules definition
	
	    f.setStrategy(Strategy.PARTIAL_AGGREGATION);
	    f.setStatisticExtractor(SwExtractorByBin.class);
	    f.setStatisticAggregator(SwAggregatorByBin.class);
	    f.addAFFunctionEvaluator(FSWM.class);
	
	    # Step 3: Pipeline execution
	
	    f.compute();
	    f.save();
	    f.close();
        }
}
	\end{lstlisting}
\end{minipage}

\begin{minipage}{\linewidth}
	\begin{lstlisting}[frame = single, caption={Configuration file equivalent to the code reported in Listing \ref{lst:sw}.}, label=lst:sw_conf, stepnumber=1, xleftmargin=15pt, xrightmargin=20pt]

# Input, output and parameters definition
pattern=100101000100011001
k=18
threshold=0
slices=2048
input=data/*.fasta
output=distances
task=distance

# Modules definition
strategy=partial_aggregation
extractor=fade.sw.SwExtractorByBin         
aggregator=fade.sw.SwAggregatorByBin           
evaluator=fade.affunction.FSWM 
\end{lstlisting}
\end{minipage}

\subsubsection{A General Paradigm}

Our framework introduces a general paradigm for building AF  analysis pipelines over a collection of input sequences, independently of the statistics being collected and of the AF function being used. Assuming that the required modules are already available in our standard primitives library or have been provided using the procedure described in Section \ref{sec:PE}, the paradigm is comprised of the following steps:

\begin{enumerate}
	\item{\bf Input, output and parameters definition:} the location of the input and of the output files is provided, as well as the definition of the parameters required by the pipeline modules.
	\item{\bf Modules definition: } the modules to use for the different stages of the pipeline are defined. The modules to be used in Stage 2 and in Stage 3 to filter partial and aggregated statistics are optional (i.e., if no filtering is required). More modules can be defined for Stage 5 (i.e., if two or more AF functions have be to evaluated).
	\item{\bf Pipeline execution:} the assembled pipeline is run over the sequences found in the input location. The AF matrices returned by the pipeline can be either saved on the output location or used, as input, for further processing. 
\end{enumerate}



\subsection{Possible Extensions}\label{sec:PE}

Coherently with the architecture of our framework reported in Figure \ref{fig:architecture}, it is possible to add support for more statistics and/or AF functions by properly deriving and specializing some standard classes, as described next.

\subsubsection{Supporting More Statistics.}
The user can add support for a target statistic not originally included in the library by extending and specializing a set of standard classes in the following way:
\begin{itemize}
    \item Inherit and specialize the {\tt Statistic} class to provide a Java representation for the target statistic.
    \item Inherit and specialize the {\tt Value} class to provide a Java representation for the partial or aggregated value assumed by the target statistic.
    \item Inherit and specialize the {\tt Data} class to provide a Java representation for a collection of occurrences of the target statistic.
    \item Inherit and specialize the {\tt StatisticExtractor} class to provide a method to be used for extracting occurrences of the target statistic or collection of occurrences of the target statistic from an input sequence. From one to three versions of this method can be provided according to the partitioning strategy to support (see Section \ref{partitioningstrategies}).
    \item Inherit and specialize the {\tt StatisticAggregator} class to provide a method to be used for aggregating occurrences of the target statistic or collection of occurrences of the target statistic. From one to three versions of this method can be provided according to the partitioning strategy to support (see Section \ref{partitioningstrategies}).



\end{itemize}


\subsubsection{Supporting More AF Functions.}

The user can add support for a target AF function not originally included in the library by extending and specializing the \texttt{AFFunctionEvaluator} class. This class provides a customizable implementation of a distributed algorithm for evaluating an AF function over a distributed collection $\mathcal{C}$ of statistics. The algorithm is made of the following steps:
\begin{enumerate}
    \item {\em Partial Distance/Similarity Evaluation.} Given a pair of statistics in $\mathcal{C}$, evaluates their partial distance/similarity by means of a user-provided associative and commutative two-arguments function.
    \item {\em Partial Distances/Similarities Combination.} Given a pair of partial distances/similarities, evaluates their combination by means of a user-provided associative and commutative two-arguments function.
    \item {\em Overall Distance/Similarity Finalization.} Given the overall distance/similarity resulting from the combination of all partial distances/similarities, finalizes its value by means of a user-provided function.
\end{enumerate}

Steps $1$ and $2$ of the algorithm are run in parallel on the different nodes of the distributed system holding a part of $\mathcal{C}$. Once each node has processed all of its local statistics, the resulting partial distance/similarity is combined with those of other nodes by means of step $2$. The algorithm ends with the execution of step $3$ on the value resulting from the combination of all partial distances/similarities.

An example of implementation for the \textit{Euclidean} distance is shown in Listing \ref{lst:euclidean}. Here the \texttt{evaluatePar-} \texttt{tialAFValue} method is used to evaluate the partial euclidean distance, given a pair of statistics in input, as the square of the difference of their counts.  The \texttt{combinePartialAFValues} method is used to combine a pair of partial distances/similarities as a unique distance/similarity value using the sum operator. Finally, the \texttt{finalizeAFValue} method is used to finalize the computed distance/similarity, by evaluating the square root over the sum of the square of the differences of all statistics.



\begin{minipage}{\linewidth}

\begin{lstlisting}[language=Java, frame=single, caption=Java code of the class implementing support for the Euclidean Distance in our framework, label=lst:euclidean, numbers=left, xleftmargin=15pt, xrightmargin=20pt]
public class Euclidean extends AFFunctionEvaluatorByStatistic {
    public AFValue evaluatePartialAFValue(Value v1, Value v2) {
		long count1 = ((CountValue)v1).count;
		long count2 = ((CountValue)v2).count;
    
        return new AFValue(Math.pow(count1 - count2, 2));
    }
    
    public AFValue combinePartialAFValues(AFValue afv1, 
        AFValue afv2) {
        return new AFValue(afv1.value + afv2.value);
    }

    public AFValue finalizeAFValue(AFValue afv) {
        return new AFValue(Math.sqrt(afv.value));
    }
}
\end{lstlisting}
\end{minipage}


\begin{figure}
	\centering
	\includegraphics[scale=0.27]{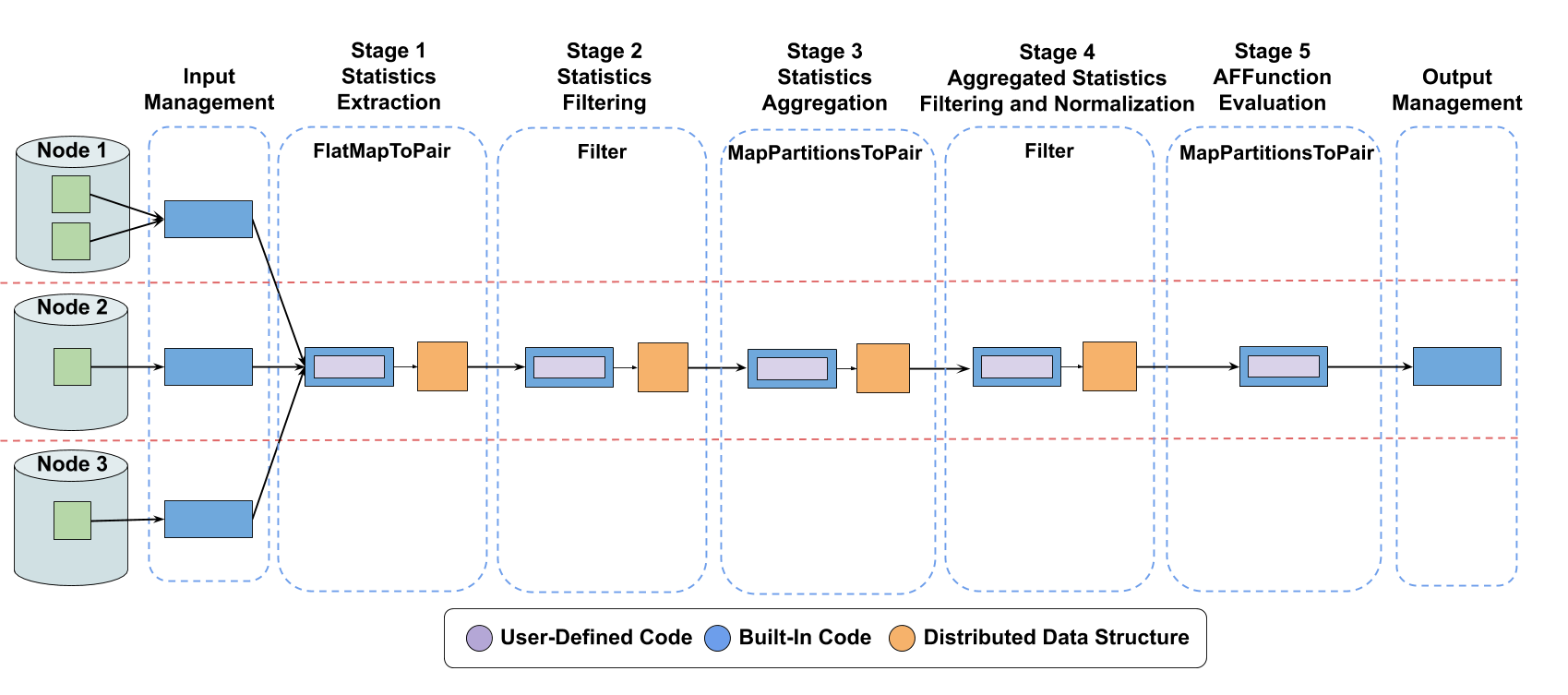}
	\caption{{\bf Total Aggregation Strategy: } layout of the basic pipeline available in our framework when used to process a collection of input sequences on three different nodes of a distributed system by means of the total aggregation partitioning strategy. }
	\label{fig:total_aggr}
\end{figure}

\begin{figure}
	\centering
	\includegraphics[scale=0.27]{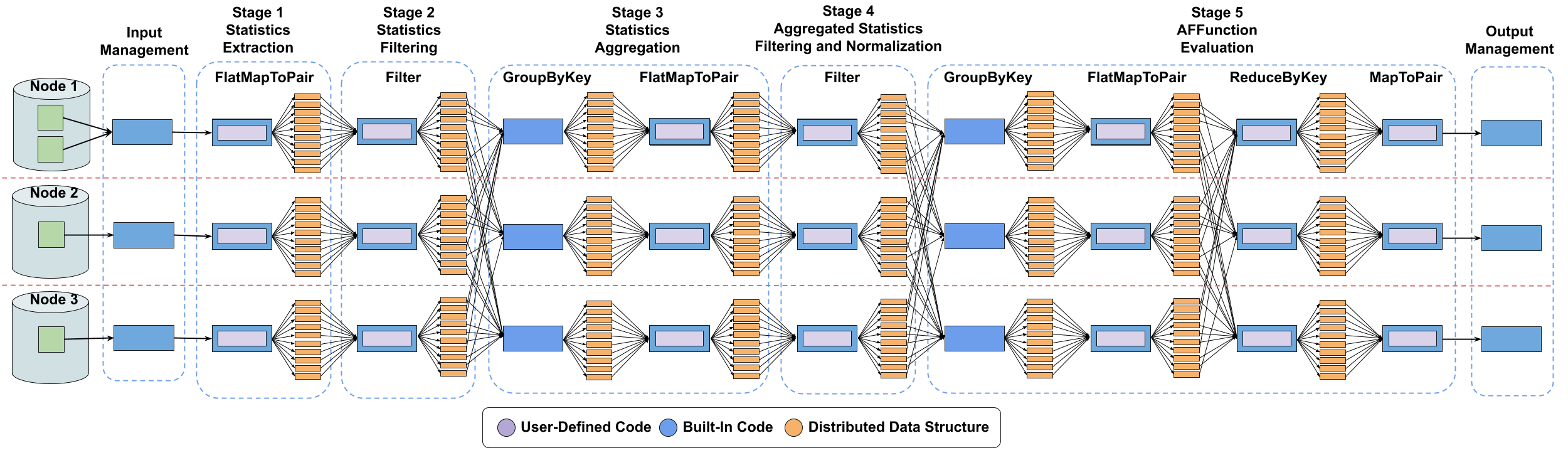}
	\caption{{\bf No Aggregation Strategy: } layout of the basic pipeline available in our framework when used to process a collection of input sequences on three different nodes of a distributed system by means of the no aggregation partitioning strategy. }
	\label{fig:no_aggr}
\end{figure}

\begin{figure}
	\centering
	\includegraphics[scale=0.27]{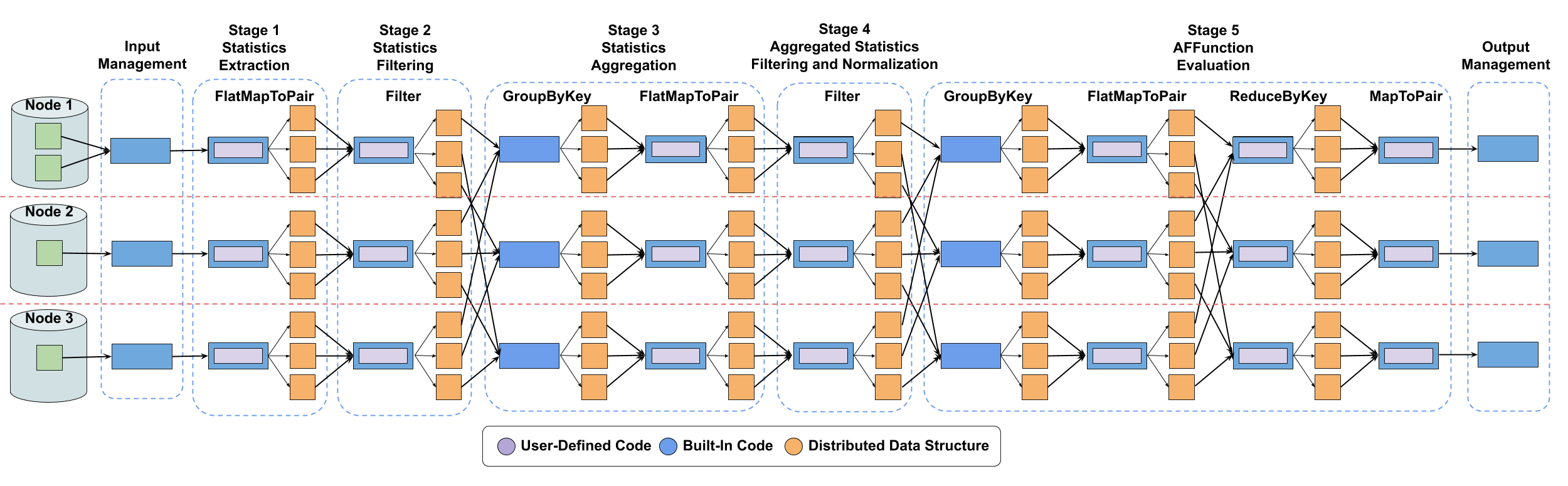}
	\caption{{\bf Partial Aggregation Strategy: }  layout of the basic pipeline available in our framework when used to process a collection of input sequences on three different nodes of a distributed system by means of the partial aggregation partitioning strategy. }
	\label{fig:part_aggr}
\end{figure}

\section{Architectural Engineering: Details}\label{sec:AED}
\label{partitioningstrategies}




The three different data partitioning strategies implemented by our basic pipelines are briefly introduced and discussed. Intuitively, they transform the basic logical pipeline introduced in Figure 1 of the \MM into one of the pipelines exemplified in Figures \ref{fig:total_aggr}-\ref{fig:part_aggr} for three nodes. The pipeline selected by the user via the choice of the aggregation strategy is then executed. The transformation from logical  to \vir{physical} is transparent to the user. 

\paragraph{Strategy 1: Total Aggregation.} 



This strategy implements the basic logical pipeline introduced in Figure 1 of the \MM as the physical pipeline reported in Figure \ref{fig:total_aggr}.  Among the proposed strategies, this is the one more closely resembling  the logical pipeline as all stages are executed in a sequential way, by aggregating all data records on a single node of the distributed system.
In the following list, details are provided about the way each stage of the basic pipeline is implemented by this strategy. 

\begin{itemize}
	\item \texttt{StatisticExtractor.} It takes a genomic sequence, or a chunk of it, as input, and returns a single list with an associated value. The resulting distributed data structure has the form \KV{unique}{(idSeq, V)}, where \emph{unique} is the id used to group all the records into a single node, \emph{idSeq} is the id of the sequence where the statistic comes from and \emph{V} is its partial record. This data structure contains, at least, a single record associated to the statistics extracted from all the tasks and from all the sequences. This stage is implemented as a Spark \texttt{flatMapToPair} transformation.
	\item \texttt{StatisticFilter.} Filters from the list of partial statistics the ones satisfying a provided boolean condition. It is implemented as a Spark \texttt{filter} transformation. 
	
	\item \texttt{StatisticAggregator.} Combines all partial statistics according to a user-provided function. The resulting distributed data structure has the form \KV{unique}{(idSeq, S, V)}, where \emph{unique} is the id used to group all the records into a single node, \emph{S} is the extracted statistic, \emph{idSeq} is the id of the sequence where \emph{S} comes from, and \emph{V} is its aggregated value.  It is implemented as a Spark \texttt{mapPartitionsToPair} transformation.
\item \texttt{AggregatedStatisticFilter.} Filters from the list of aggregated statistics the ones satisfying a provided boolean condition. It is implemented as a Spark \texttt{filter} transformation. 
	\item \texttt{AFFunctionEvaluator.} Given the statistics aggregations, it computes the partial distances/similarities between all the pairs of sequences. The partial distances/similarities are then combined and, if necessary, finalized. It is implemented as a Spark \texttt{mapPartitionsToPair} transformation.
\end{itemize}

\paragraph{Strategy 2: No Aggregation.} 

This strategy implements the logical pipeline introduced in Figure 1 of the \MM as the physical pipeline reported in Figure \ref{fig:no_aggr}. Differently from the previous strategy, this one aims at maximizing the degree of parallelism as each data record is processed independently of the others in a separate task. In the following list, details are provided about the way each stage of the basic pipeline is implemented by this strategy. 

\begin{itemize}
	\item \texttt{StatisticExtractor.} It takes a genomic sequence, or a chunk of it, as input, and returns a list of partial statistics with an associated value. Each partial statistic is encoded as a stand-alone data record, distinct from the others. The resulting distributed data structure has the form \KV{S}{(idSeq,V)}, where \emph{S} is the extracted statistic, \emph{idSeq} is the id of the sequence where \emph{S} comes from, and \emph{V} is its partial value. At the end of the stage, this data structure contains, at least, as many records as the number of distinct statistics extracted from all the tasks and from all the sequences.  This stage uses a Spark \texttt{flatMapToPair} transformation. 
	\item \texttt{StatisticFilter.} As in Strategy 1.
\item \texttt{StatisticAggregator.} Groups on a node all the occurrences of a same statistic for the same sequence, then combines them according to a customizable function. The resulting distributed data structure has the form \KV{S}{(idSeq,V)}, where \emph{S} is the extracted statistic, \emph{idSeq} is the id of the sequence where \emph{S} comes from, and \emph{V} is its aggregated value.  This stage uses a Spark \texttt{groupByKey} and a Spark \texttt{flatMapToPair} transformations. 
	\item \texttt{AggregatedStatisticFilter.} As in Strategy 1.
	
\item \texttt{AFFunctionEvaluator.} For each of the aggregated statistics resulting from the previous stage, it computes the partial distances/similarities between all the pairs of sequences. Then, for each pair of sequences, the resulting overall distance/similarity is obtained by combining their partial distances/similarities. It is implemented using, respectively, a Spark \texttt{groupByKey} transformation (used to gather on a same node all the occurrences of a same aggregated statistic), a Spark \texttt{flatMapToPair} transformation (used to compute the partial distances/similarities), a Spark \texttt{reduceByKey} transformation (used to combine the partial distances/similarities) and a Spark \texttt{mapToPair} transformation (used to finalize the distances/similarities).
\end{itemize}

\paragraph{Strategy 3: Partial Aggregation.}

 
 This strategy implements the logical pipeline introduced in Figure 1 of the \MM as the physical pipeline reported in Figure \ref{fig:part_aggr}. It provides a sort of balance between the previous two strategies as it allows to improve performance by processing large data aggregations in a distributed way.

 In the following list, details are provided about the way each stage of the basic pipeline is implemented by this strategy. 


\begin{itemize}
	\item \texttt{StatisticExtractor.} It takes a genomic sequence, or a chunk of it, as input, and returns a list of statistics with an associated value and partitioned into bins. The resulting distributed data structure has the form \KV{idBin}{(idSeq,V)}, where \emph{idBin} is the id of the bin, \emph{idSeq} is the id of the sequence where the statistics comes from and \emph{V} is a set of statistics with partial values. It is possible to customize the size of this data structure by modifying the number of the bins used for the partitioning.  This stage uses a Spark \texttt{flatMapToPair} transformation.
	\item \texttt{StatisticFilter.} As in Strategy 1.

	\item \texttt{StatisticAggregator.} Groups on a node all the partial statistics stored in a same bin, then combines them according to the sequence they come from and using a user-provided function. The resulting distributed data structure has the form  \KV{idBin}{(idSeq,S,V)}, where \emph{idBin} is the id of the bin, \emph{idSeq} is the id of the sequence where the statistics comes from, \emph{S} is the aggregated statistic and  \emph{V} is its value. This stage uses a Spark \texttt{groupByKey} and \texttt{flatMapToPair} transformations. 
	\item \texttt{AggregatedStatisticFilter.} As in Strategy 1.
	\item \texttt{AFFunctionEvaluator.} Given the statistics aggregations, it computes the partial distances/similarities between all the pairs of sequences. The partial distances/similarities are first gathered according to the bin they belong to, then are combined and, if necessary, finalized. It is implemented as Spark \texttt{groupByKey} transformation, \texttt{flatMapToPair} transformation (used to compute the partial distances/similarities), \texttt{reduceByKey} transformation (used to combine the partial distances/similarities), \texttt{mapToPair} transformation (used to finalize the distances/similarities).
\end{itemize}


\section{A Test of  Consistent Significance  of an AF Function: Spark Implementation Details}
\label{sec:corruption}

\begin{figure}[!ht]
	\centering
	\includegraphics[width=0.8\textwidth]{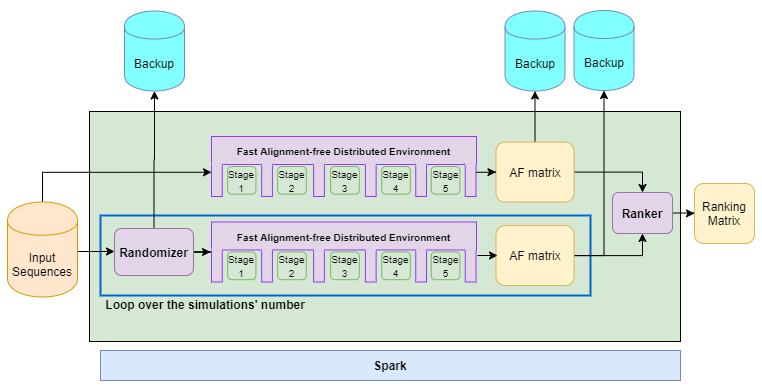}
	\caption{A layout of the architecture of the pipeline for the fast hypothesis testing of alignment-free algorithms. It repeatedly uses, as subroutine, the basic pipeline for the fast computation of AF functions. Backup primitives are used to save the progresses of a test so as to recover an interrupted computation or to allow a very long execution to be broken in shorter parts. }
	\label{simulation}
\end{figure}

With reference to the Hypothesis Test method outlined in Section 2.5.1 of the \MM, its Spark modules are summarized in Figure \ref{simulation}. The upper pipeline processes the set $S$, while repeated executions of the lower pipeline execute the simulation. The module {\bf Randomizer} generates in each run the random datasets and it is executed in a distributed way over the architecture.  Finally, for each entry of the AF matrix being tested, a ranking is produced accounting for the corresponding entry in the simulated AF matrices. The module {\bf Ranker} is responsible for that process. Once again, it is executed in a distributed way over the architecture. The final result is a matrix, where each entry contains the rank of the original entry with respect to the simulated ones. The desired confidence level and Bonferroni correction determine which entries pass the test, i.e., for which the null hypothesis can be rejected,  and  whether the AF matrix passes the test as a whole.

\ignore{
The datasets used during the experiments are the following:
\begin{itemize}
\Comment{
	\item \textit{CRM (small)}. The \textit{CRM} dataset pertains to cis-regulatory sequence modules (CRMs) that are known to regulate expression in the same tissue and/or development stage in fly or human. A CRM can be loosely defined as a contiguous non-coding sequence that contains multiple transcription factor binding sites and drives some aspect of a gene's expression profile. It was originally collected by Kantorovitz {\em et al.} \citep{kantorovitz2007statistical} to test the capacity of AF methods of identifying functional relationships between regulatory sequences.

The dataset is a FASTA file containing a mixture of 7 subsets of CRM sequences, each taken from different tissues of {\em D. melanogaster} or {\em Homo sapiens}. Each of the 7 subsets has $2\dot n$ sequences, where the first $n$ sequences are CRMs ("positive half") and the remaining ones are random non-coding sequences with matching lengths, chosen from the respective genome ("negative half"). 
}
	\item \textit{Microbial (medium) \cite{metaref2013}}.

	\item \textit{Plants (large) \cite{AFproject2019}}. The \textit{Plants} dataset pertains to full genome sequences of $14$ plant species. It is organized as a collection of $14$ FASTA files and it was originally compiled by Hatje and Kollmar and presented in \citep{hatje2012phylogenetic}.
\end{itemize}
}

%

\ignore{
The dataset is organized as a collection of $6$ FASTQ files containing raw sequence data: three of these files refers to microbiomes sampled from the stool body set and three refers to the tongue dorsum bodyset.

}

\section{Datasets and Hardware.}

\label{subsec:datasets}

The datasets used in our study are:

\begin{itemize}
	\item {\bf E.coli/Shigella.} It contains $27$ genomic sequences of the {\em Bacteria} taxonomic group, having an average length of $4,905,896$ bp. For further details, see \cite{Bernard16,Zielezinski2019}. 
	
	\item {\bf Mithocondria.}  It contains a collection of $25$ different genomes of fish species of the suborder \textit{Labroidei}, having an average size of $16,618$ bp. For further details, see \cite{Fischer13,Zielezinski2019}. 	
	
	\item {\bf Plants.} It contains $14$ assembled very-large genomes of the {\em Plants} taxonomic group, having an average size of $337,515,688$ bp each. For further details, see \cite{Zielezinski2019}. 
	
	\item {\bf Unassembled E.coli strains.}  The dataset is the unassembled version of the {\bf E.coli strains} taxonomic group, with an average read length of $150$ bp. The sequencing coverages considered are: 0.03125 ($29,557$ reads), 0.125 ($118,266$ reads), 1 ($946,169$ reads). For further details, see \cite{yi2013co,Zielezinski2019}. 
	
	\item {\bf Unassembled Plants.}  The dataset is the unassembled version of the {\bf Plants} taxonomic group, with an average read length of $150$ bp. The sequencing coverages considered are: 1 ($30,903,727$ reads). For further details, see \cite{Zielezinski2019}. 
	
	\item {\bf Yersinia.}  It contains $8$ genomic sequences of the {\em Bacteria} taxonomic group. The average sequence length is $4,605,552$ bp. For further details, see \cite{Bernard16,Zielezinski2019}. 
\end{itemize}

\paragraph{Hardware Platform.}
All the experiments have been performed on a $25$ nodes Linux-based cluster, with one node acting as \textit{resource manager} and the remaining nodes being used as workers. The cluster is installed with Hadoop 2.8.1 and the Spark 2.2 software distributions. Each node of this cluster is equipped with one 8-core Intel Xeon E3-12@2.40 GHz processor and 32GB of RAM. Moreover, each node has a 200 GB virtual disk reserved to HDFS, for an overall capacity of about 6 TB.

\begin{table}[htb]
\centering
\begin{tabular}{lccccc}
\hline
& \bf{l} & \bf{k} & \bf{w }& \bf{L} & \bf{s} \\
\hline
\bf{E.coli/Shigella} & 100 & 11 & 12 & 112 & 1000\\
\bf{Mitochondria} & 100 & 7 & 12 & 112 & 1000 \\
\bf{Plants} & 80 & 14 & 12 & 112 & 1000 \\
\bf{Unass. E.coli (cov. 0.03125)} & 100 & 8 & 12 & 72 & 1000 \\
\bf{Unass. E.coli (cov. 0.125)} & 100 & 9 & 12 & 72 & 1000\\
\bf{Unass. E.coli (cov. 1)} & 100 & 11 & 12 & 72 & 1000 \\
\bf{Unass. Plants (cov. 1)} & 80 & 14 & 12 & 72 & 1000 \\
\bf{Yersinia} & 100 & 11 & 12 & 112 & 1000 \\
\hline
\end{tabular}
\caption{Outline of the parameters used in our experiment for assessing the statistical significance of several AF functions on the considered datasets. $l$ denotes the number of runs used for the Monte Carlo simulation. $k$ denotes the length of the \kmers used for running AF functions based on histogram statistics. $w$ and $L$ are respectively the length and the weight of the pattern used for running spaced words AF functions. $s$ is the size of the sketches used when running the MinHash approximation algorithm.}
\label{tab:parameters}
\end{table} 

%
%

\section{AF Functions Parameters and Monte Carlo Simulation Iterations}\label{sec:resk}

For each experiment, the number of Monte Carlo Simulations has been chosen so as to obtain the results of the test in a reasonable amount of time, given the size of the dataset. 
The criteria for the choice of relevant parameters
for AF functions is provided next, while a  summary of all the  parameters used in our experiments is available in Table \ref{tab:parameters}. 

\subsection{Histogram Statistics: choice of k}
\label{sec:choicek}

As well argued in \cite{luczak2017survey}, for histogram-based AF functions, the choice of the \kmers $k$ is crucial for the success of those methods. It is a heuristic choice. The one that seems to work best is the one shown in equation (\ref{eq:k}). 

\begin{equation}
\label{eq:k}
    k = \lceil log_{4}\left( \frac{1}{|S|}\sum_{i \in  S}{len(i)}\right) \rceil - 1
\end{equation}

\noindent in which $n$ is the total number of sequences. It is also to be remarked that the choice of $k$ should  be also in a range that preserves the \vir{information theoretic content} of the sequences to be analyzed. To the best of our knowledge, two closely related approaches  have been proposed in the literature \cite{Giancarlo2015,sims2011whole}. Here we follow the approach in \cite{Giancarlo2015}.  Then, the value of $k$ for the AF histogram-based function to be used for the analysis is chosen as the minimum between the $k'$ returned by this approach and the $k$ coming out of equation (\ref{eq:k}).

As for the choice of the sketch size to be used when running the MinHash approximation algorithm Mash, we considered the default value of $1,000$ \cite{Ondov2016}.

\subsection{Word matches: spaced word methods}
\label{sec:swsparameters}
When considering \sws methods, there are two parameters that need to be fixed: the length of the pattern \PAT and its weight $w$. In our experiments with assembled genomes,  we have used the same default values considered in \cite{Leimester17}, i.e., a pattern with $w=12$ and $100$ don't care positions. Instead, when analyzing collections of reads we resorted to the default values used by the authors of \cite{lau2019read} to analyze short sequencing reads: a pattern with $w=12$ and $60$ don't care positions.


\ignore{
\begin{table}[ht]
\centering
\begin{tabular}{|l|c|c|c|c|c|c|}
\hline
& \multicolumn{2}{c}{ \textbf{q=1} } & \multicolumn{2}{c}{ \textbf{q=7} } & \multicolumn{2}{c|}{ \textbf{q=10} } \\
& \textbf{corr.} & \textbf{pvalue} & \textbf{corr.} & \textbf{pvalue} & \textbf{corr.} & \textbf{pvalue} \\
\hline
\textbf{D2} & 9.04E-01 & \textbf{1.33E-04} & 8.85E-01 & \textbf{2.95E-04} & 8.87E-01 & \textbf{2.68E-04} \\
\textbf{D2Z} & 8.90E-01 & \textbf{2.43E-04} & 8.97E-01 & \textbf{1.82E-04} & 8.90E-01 & \textbf{2.38E-04} \\
\textbf{FSWM} & 3.96E-01 & 2.28E-01 & 3.96E-01 & 2.28E-01 & 3.96E-01 & 2.28E-01 \\
\hline
\end{tabular}
\caption{Correlation values between the percentage of noisy entries over the AF matrix obtained by running several AF functions over the Mitochondria dataset while using different values of $q$ and the RF score of these matrices. The percentage of noisy entries has a range that goes from $0\%$ to $100\%$ with a $10\%$ step. In bold the cases that are statistically significant, with p-value<=5\%.}
\label{tab:corr_mito}
\end{table}

\begin{table}[ht]
\centering
\begin{tabular}{|l|c|c|c|c|c|c|}
\hline
& \multicolumn{2}{c}{ \textbf{q=1} } & \multicolumn{2}{c}{ \textbf{q=7} } & \multicolumn{2}{c|}{ \textbf{q=10} } \\
& \textbf{corr.} & \textbf{pvalue} & \textbf{corr.} & \textbf{pvalue} & \textbf{corr.} & \textbf{pvalue} \\
\hline
\textbf{D2} & 5.42E-01 & 8.53E-02 & 6.07E-01 & \textbf{4.77E-02} & 6.07E-01 & \textbf{4.77E-02} \\
\textbf{D2Z} &  6.67E-01 & \textbf{2.49E-02} & 8.45E-01 & \textbf{1.06E-03} & 8.65E-01 & \textbf{5.89E-04} \\
\textbf{FSWM} & 5.80E-01 & 6.15E-02 & 0 & 1 & 3.46E-01 & 2.97E-01 \\
\textbf{D2s} & 6.72E-01 & \textbf{2.37E-02} & 6.72E-01 & \textbf{2.37E-02} & 7.36E-01 & \textbf{9.88E-03}\\
\textbf{D2*} & 6.44E-01 & \textbf{3.24E-02} & 6.44E-01 & \textbf{3.24E-02} & 8.37E-01 & \textbf{1.32E-03}\\
\hline
\end{tabular}
\caption{Correlation values between the percentage of noisy entries over the AF matrix obtained by running several AF functions over the Yersinia dataset while using different values of $q$ and the RF score of these matrices. The percentage of noisy entries has a range that goes from $0\%$ to $100\%$ with a $10\%$ step. In bold the cases that are statistically significant, with p-value<=5\%.}
\label{tab:corr_yers}
\end{table}

\begin{table}[ht]
\centering
\begin{tabular}{|l|c|c|c|c|c|c|}
\hline
& \multicolumn{2}{c}{ \textbf{q=1} } & \multicolumn{2}{c}{ \textbf{q=7} } & \multicolumn{2}{c|}{ \textbf{q=10} } \\
& \textbf{corr.} & \textbf{pvalue} & \textbf{corr.} & \textbf{pvalue} & \textbf{corr.} & \textbf{pvalue} \\
\hline
\textbf{D2} & 7.03E-01 & \textbf{1.58E-02} & 7.14E-01 & \textbf{1.37E-02} & 7.37E-01 & \textbf{9.60E-03} \\
\textbf{D2Z} & 8.56E-01 & \textbf{7.72E-04} & 9.28E-01 & \textbf{3.80E-05} & 9.52E-01 & \textbf{6.00E-06} \\
\textbf{FSWM} & 2.70E-01 & 4.23E-01 & 2.80E-01 & 4.04E-01 & 3.20E-01 & 3.38E-01 \\
\textbf{D2s} & 9.04E-01 & \textbf{1.32E-04} & 9.43E-01 & \textbf{1.40E-05} & 9.49E-01 & \textbf{8.00E-06}\\
\textbf{D2*} & 7.52E-01 & \textbf{7.65E-03} & 9.42E-01 & \textbf{1.50E-05} & 9.42E-01 & \textbf{1.40E-05}\\
\hline
\end{tabular}
\caption{Correlation values between the percentage of noisy entries over the AF matrix obtained by running several AF functions over the E.coli/Shigella dataset while using different values of $q$ and the RF score of these matrices. The percentage of noisy entries has a range that goes from $0\%$ to $100\%$ with a $10\%$ step. In bold the cases that are statistically significant, with p-value<=5\%.}
\label{tab:corr_shige}
\end{table}

} 

\section{Additional Figures}

\begin{figure}
	\centering
	\includegraphics[width=400pt]{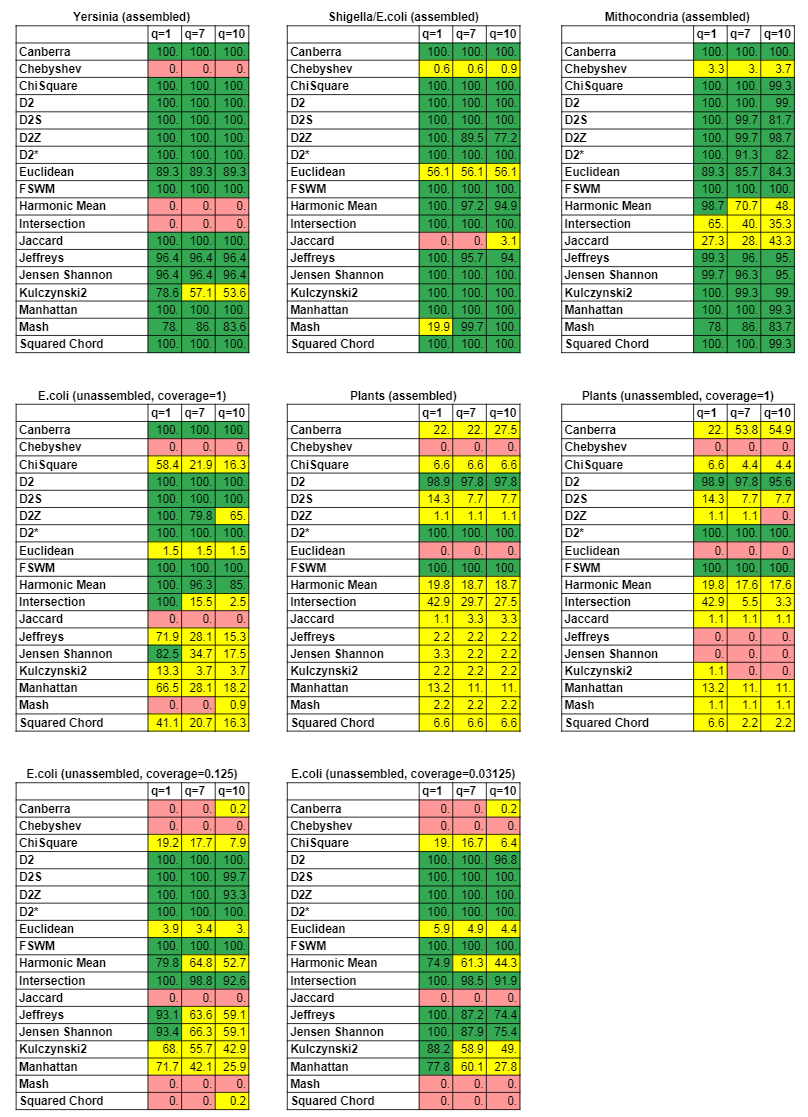}
	\caption{
Hypothesis Test complete results for the different AF functions considered in this research  when executed on three different datasets with $q={1,7,10}$ and with significance level set to $1\%$.}
	\label{fig:sigtestallall}
\end{figure}
\ignore{

\begin{figure}[ht]
	\centering
	\includegraphics[width=400pt]{significance-test-d2}
	\caption{Summary of the hypothesis test results for the D2 family functions considered in this paper when executed on different datasets with $q={1,7,10}$ and with significance level set to $1\%$. Each table reports the percentage of rejected null test hypotheses for each D2 family function and each value of q on each dataset. \label{fig:sigtest-d2}}
\end{figure}

\begin{figure}[t]
	\centering
	\includegraphics[width=\linewidth]{yersinia_sigtest}
	\caption{Ranking hypothesis test results for the different AF functions considered in this paper run on the Yersinia dataset with $q=1$ and $100$ executions. Each matrix refers to a different AF function. For each matrix, its entries report the rank of the distance evaluated by the  algorithm on the original dataset with respect to all considered datasets. On the $x$ and on the $y$ axis the input sequence identifier. Green means that the test has been passed for a particular entry $(i,j)$, red means it has not been passed. \label{fig:yersinia_sigtest}}
\end{figure}

\begin{figure}
	\centering
	\includegraphics[width=\linewidth]{plants_sigtest}
	\caption{Ranking hypothesis test results for the different AF functions considered in this paper run on the Plants dataset with $q=1$ and $80$ executions. Each matrix refers to a different AF function. For each matrix, its entries report the rank of the distance evaluated by the  algorithm on the original dataset with respect to all considered datasets. On the $x$ and on the $y$ axis the input sequence identifier. Green means that the test has been passed for a particular entry $(i,j)$, red means it has not been passed. \label{fig:plants_sigtest}}
\end{figure}
}


\begin{figure}
	\centering
	\caption{Robinson Fould corruption graph of Mitochondria dataset, varying AF function and q. Each graph reports the distance of the philogenetic trees, obtained using the Neighbor-joining and the UPGMA methods, from the gold standard, measured using the Robinson Fould metric, for the considered AF function and q, and as the number of  noisy entries grows.
}
	\label{fig:corruption-mito-rf}
	\includegraphics[scale=0.45]{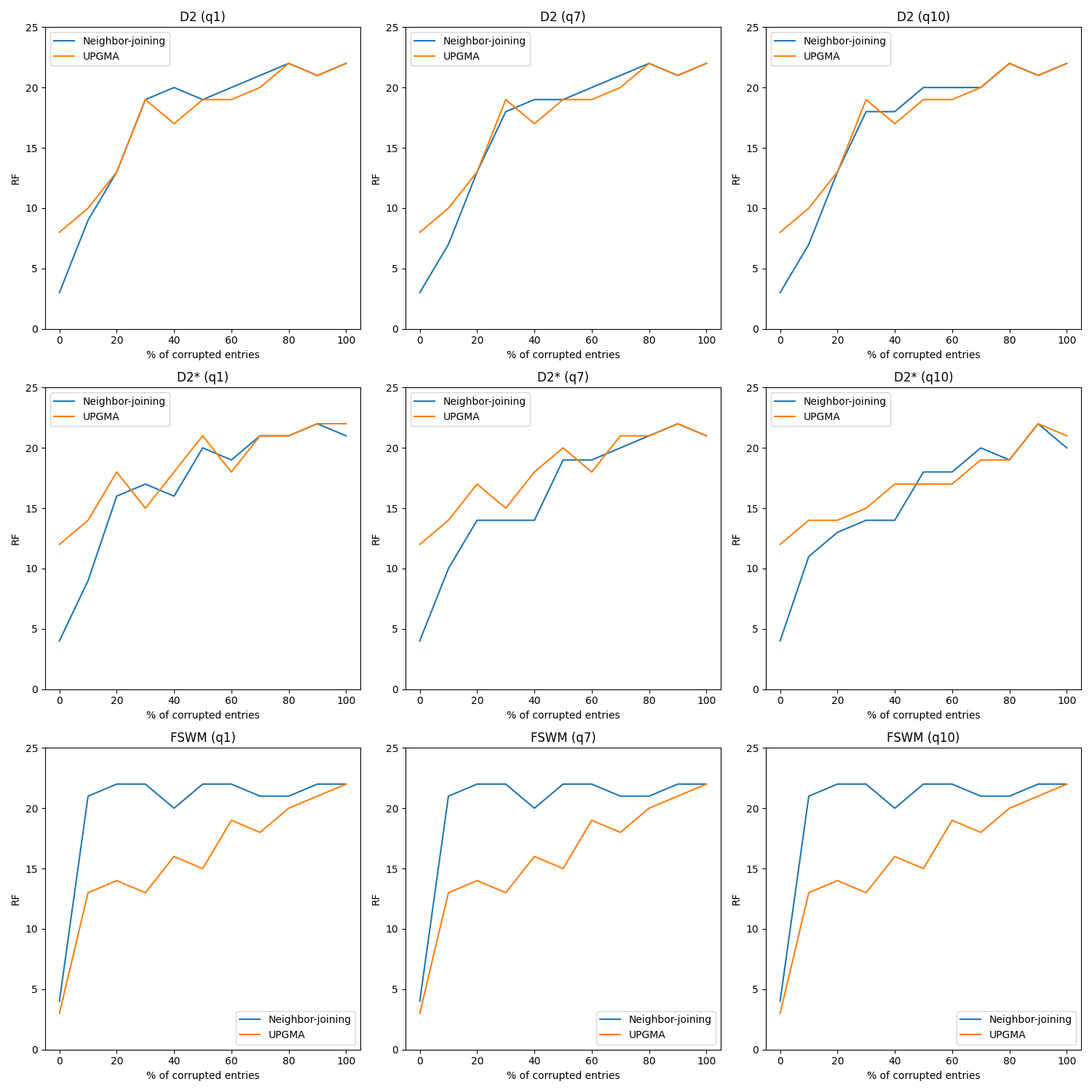}
\end{figure}

\begin{figure}
	\centering
	\caption{Robinson Fould corruption graph of Shigella dataset, varying AF function and q. Each graph reports the distance of the philogenetic trees, obtained using the Neighbor-joining and the UPGMA methods, from the gold standard, measured using the Robinson Fould metric, for the considered AF function and q, and as the number of  noisy entries grows.}
	\label{fig:corruption-shigella-rf}
	\includegraphics[scale=0.45]{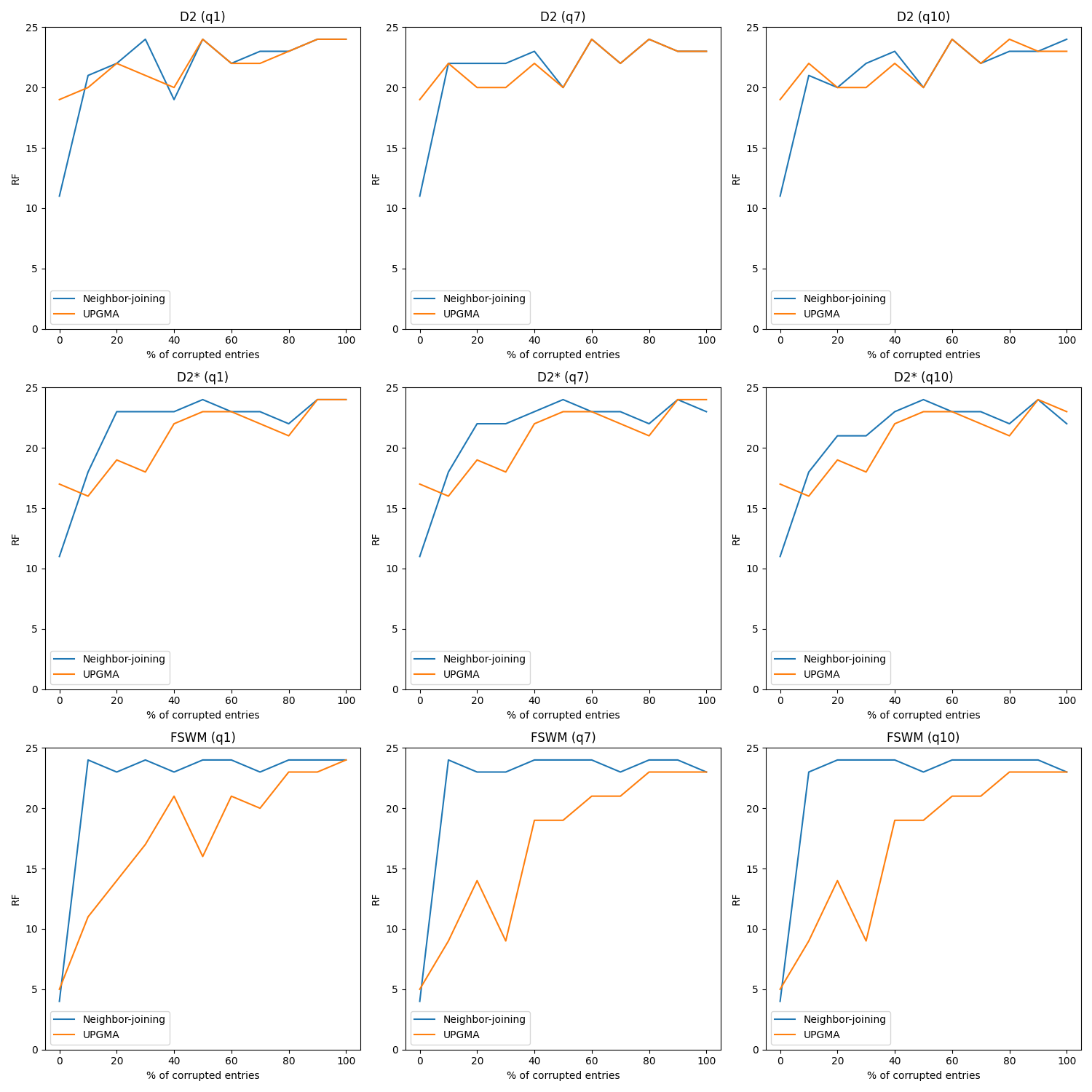}
\end{figure}

\begin{figure}
	\centering
	\caption{Robinson Fould corruption graph of Yersinia dataset, varying AF function and q. Each graph reports the distance of the philogenetic trees, obtained using the Neighbor-joining and the UPGMA methods, from the gold standard, measured using the Robinson Fould metric, for the considered AF function and q, and as the number of  noisy entries grows.}
	\label{fig:corruption-yersinia-rf}
	\includegraphics[scale=0.45]{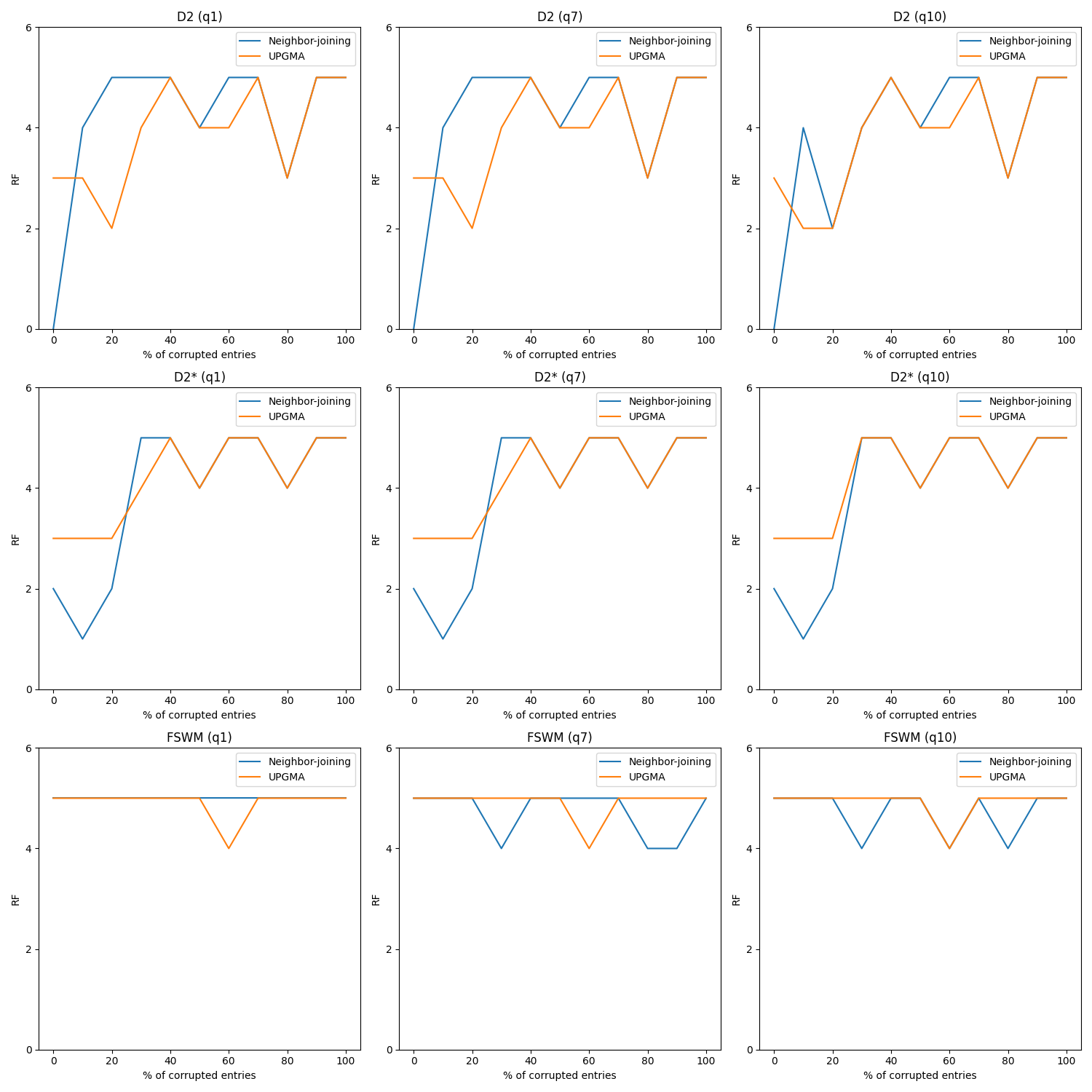}
\end{figure}

\begin{figure}
	\centering
	\caption{Matching Cluster metric corruption graph of Mitochondria dataset, varying AF function and q. Each graph reports the distance of the philogenetic trees, obtained using the Neighbor-joining and the UPGMA methods, from the gold standard, measured using the MCM metric, for the considered AF function and q, and as the number of  noisy entries grows.}
	\label{fig:corruption-mito-mcm}
	\includegraphics[scale=0.45]{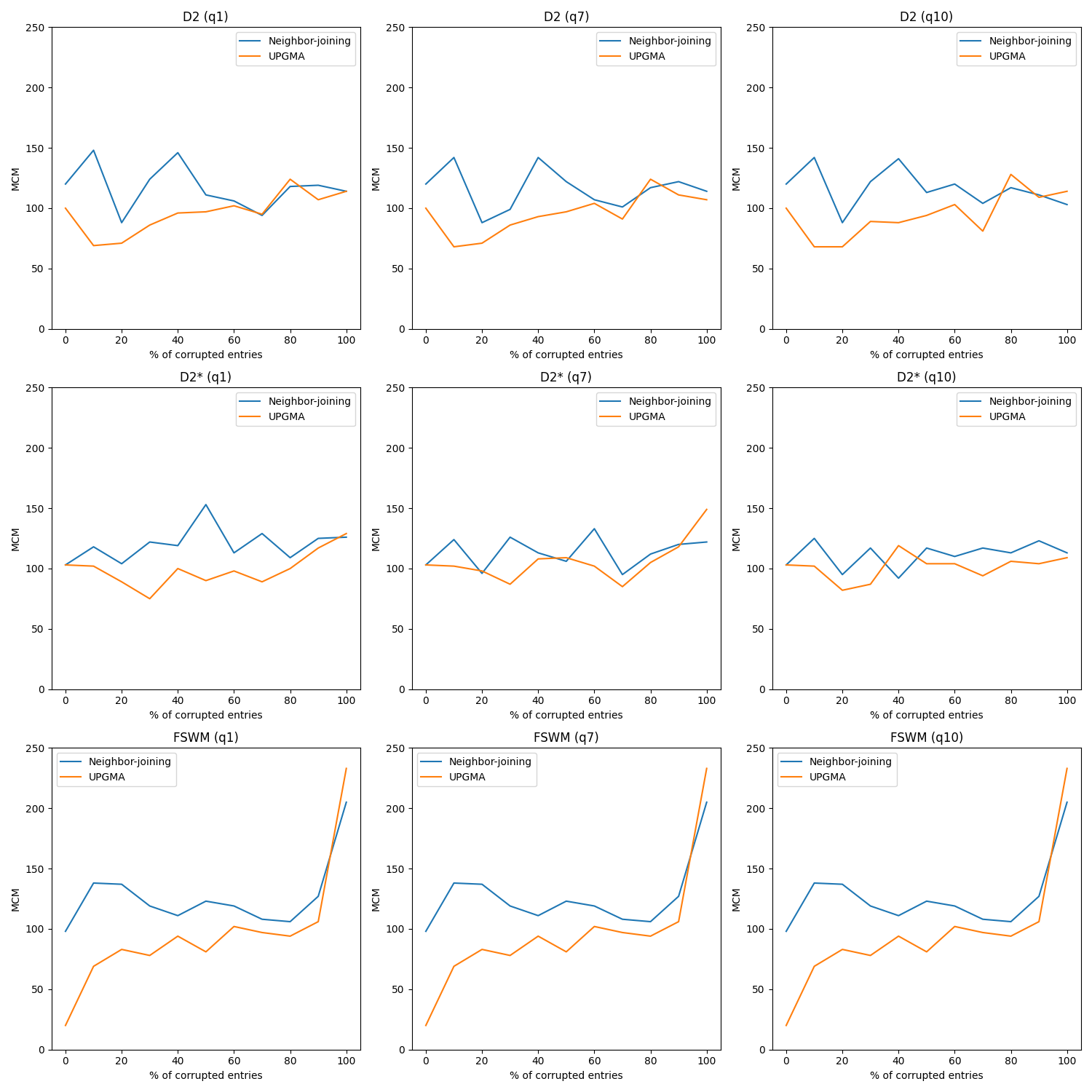}
\end{figure}

\begin{figure}
	\centering
	\caption{Matching Cluster metric corruption graph of Shigella dataset, varying AF function and q. Each graph reports the distance of the philogenetic trees, obtained using the Neighbor-joining and the UPGMA methods, from the gold standard, measured using the MCM metric, for the considered AF function and q, and as the number of  noisy entries grows.}
	\label{fig:corruption-shigella-mcm}
	\includegraphics[scale=0.45]{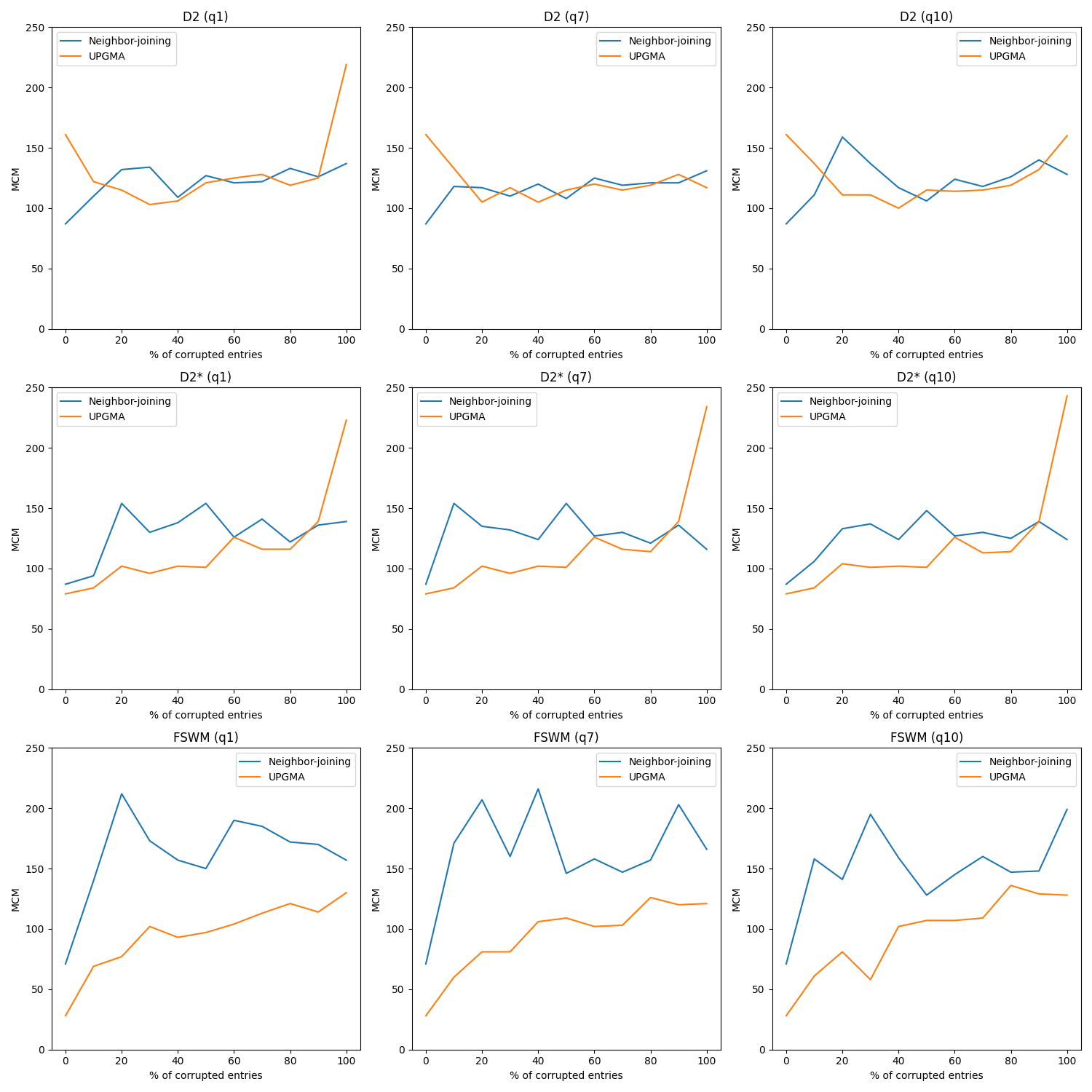}
\end{figure}

\begin{figure}
	\centering
	\caption{Matching Cluster metric corruption graph of Yersinia dataset, varying AF function and q. Each graph reports the distance of the philogenetic trees, obtained using the Neighbor-joining and the UPGMA methods, from the gold standard, measured using the MCM metric, for the considered AF function and q, and as the number of  noisy entries grows.}
	\label{fig:corruption-yersinia-mcm}
	\includegraphics[scale=0.45]{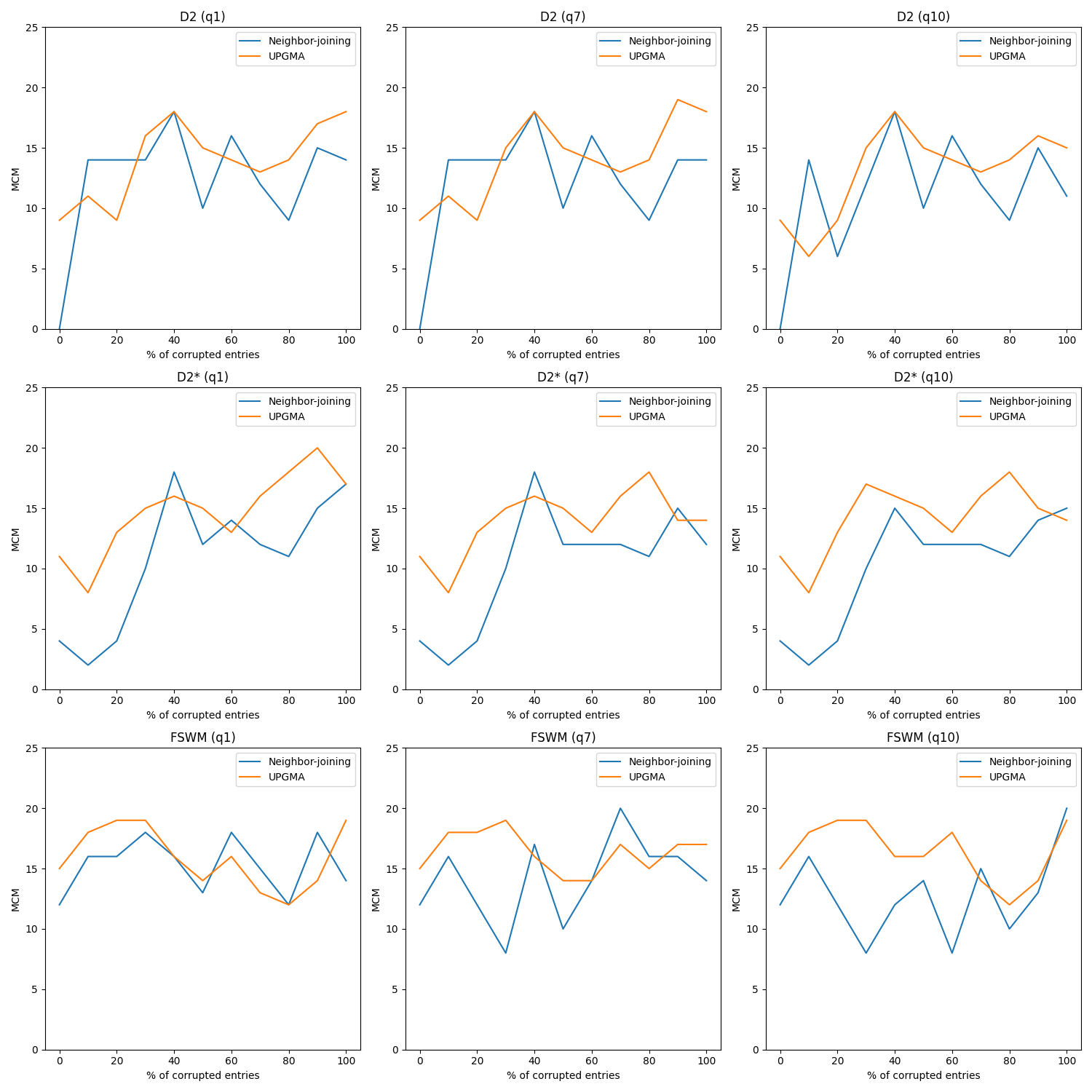}
\end{figure}

\clearpage

\clearpage

\bibliographystyle{abbrv}
\bibliography{supplementary}